\documentclass[aps,pre,showpacs,noshowkeys,amsmath,amssymb,amsfonts,superscriptaddress,longbibliography,reprint]{revtex4-1}
\usepackage[english]{babel}

\usepackage{graphicx}
\usepackage{bm}
\usepackage{physics}
\usepackage{mathtools}
\usepackage{gensymb}

\DeclareMathOperator{\arctanh}{arctanh}
\setcitestyle{super}
\usepackage{caption}
\usepackage{subcaption}
\DeclareCaptionLabelSeparator{bar}{~\rule[-0.4ex]{0.2ex}{1em}~}
\DeclareCaptionLabelFormat{subfor}{\textbf{#2}}
\newcommand*\bfcaption[2]{\caption[#1]{\textbf{#1.}#2}}
\usepackage{xcolor}
\definecolor{UBcolor}{HTML}{007CC1}
\usepackage[colorlinks=true,pdfnewwindow=true,linkcolor=UBcolor,citecolor=UBcolor,urlcolor=UBcolor,breaklinks=true,linktocpage]{hyperref}
\usepackage[all]{hypcap}
\usepackage[nameinlink,capitalise]{cleveref}
\crefname{SI section}{SI Section}{SI Sections}
\Crefname{SI section}{SI Section}{SI Sections}
\begin{document}

\title{Active phase separation by turning toward regions of higher density}

\author{Jie Zhang}
\altaffiliation{These authors contributed equally to this work.}
\affiliation{Department of Physics, University of California at Santa Barbara, Santa Barbara, CA 93106, USA}

\author{Ricard Alert}
\altaffiliation{These authors contributed equally to this work.}
\affiliation{Lewis-Sigler Institute for Integrative Genomics, Princeton University, Princeton, NJ 08544, USA}
\affiliation{Princeton Center for Theoretical Science, Princeton University, Princeton, NJ 08544, USA}

\author{Jing Yan}
\affiliation{Department of Molecular, Cellular and Developmental Biology, Yale University, New Haven, CT 06511, USA}

\author{Ned S. Wingreen}
\affiliation{Lewis-Sigler Institute for Integrative Genomics, Princeton University, Princeton, NJ 08544, USA}
\affiliation{Department of Molecular Biology, Princeton University, Princeton, NJ 08544, USA}

\author{Steve Granick}
\affiliation{Center for Soft and Living Matter, Institute for Basic Science (IBS), Ulsan 44919, South Korea}
\affiliation{Departments of Chemistry and Physics, Ulsan National Institute of Science and Technology (UNIST), Ulsan 44919, South Korea}

\date{\today}

\begin{abstract}
Studies of active matter, from molecular assemblies to animal groups, have revealed two broad classes of behavior: a tendency to align yields orientational order and collective motion, whereas particle repulsion leads to self-trapping and motility-induced phase separation. Here, we report a third class of behavior: orientational interactions that produce active phase separation. Combining theory and experiments on self-propelled Janus colloids, we show that stronger repulsion on the rear than on the front of these particles produces non-reciprocal torques that reorient particle motion toward high-density regions. Particles thus self-propel toward crowded areas, which leads to phase separation. Clusters remain fluid and exhibit fast particle turnover, in contrast to the jammed clusters that typically arise from self-trapping, and interfaces are sufficiently wide that they span entire clusters. Overall, our work identifies a torque-based mechanism for phase separation in active fluids, and our theory predicts that these orientational interactions yield coexisting phases that lack internal orientational order.
\end{abstract}

\maketitle

We are interested here in motile (``self-propelled'') agents. As motility naturally implies direction, alignment interactions lead to collective motion, with flocking as an iconic example. When the motility direction is not coordinated, self-propelled particles are well understood to undergo motility-induced phase separation (MIPS) under certain conditions \cite{Cates2015,Gonnella2015,Marchetti2016,Zottl2016,Speck2020,Ma2020}. As originally conceived \cite{Tailleur2008}, the mechanism of MIPS is self-trapping: Lower particle speed in high-density regions, due to quorum sensing \cite{Tailleur2008,Rein2016} or even just due to repulsive particle collisions \cite{Fily2012,Redner2013,Bialke2013}, promotes continual accumulation of particles. This positive feedback leads to phase separation into a dilute gas and denser clusters. In the past decade, this scenario has been widely studied using theory and simulations \cite{Cates2015,Gonnella2015,Marchetti2016,Zottl2016,Speck2020,Ma2020}. Proposals to realize this scenario using synthetic active colloids \cite{Aranson2013,Zottl2016,Bechinger2016,Zhang2017a} indeed led to the observation of motility-dependent clustering \cite{Theurkauff2012,Palacci2013,Buttinoni2013,Ginot2018}. Eliminating the possible role of attractive interactions, recent experiments with purely repulsive colloids have confirmed full phase separation \cite{Geyer2019}, or separation interrupted by the effects of aligning interactions \cite{vanderLinden2019}.  

Here we show that torques on motile particles can induce phase separation. This finding is surprising because torques can easily prevent MIPS. For example, rod-shaped particles experience torques that favor alignment, thereby avoiding self-trapping and suppressing MIPS \cite{Shi2018,vanDamme2019,Jayaram2020,Grossmann2020,Bar2020}. Other kinds of orientational interactions, including dipolar torques \cite{Pu2017,vanderLinden2019,Liao2020} and velocity-alignment rules \cite{Farrell2012,Barre2014,Sese-Sansa2018,Bhattacherjee2019a}, can either hinder or promote standard repulsion-based MIPS. Unlike conventional MIPS, torque-based aggregation requires no density-induced slowdown, so the particles that condense into clusters retain substantial speed. As a result, clusters remain fluid, as opposed to the close-packed and jammed clusters typically obtained in repulsion-based MIPS \cite{Fily2012,Redner2013,Bialke2013,Geyer2019,vanderLinden2019}. Consequently, this alternative mechanism of phase separation has the potential to enlarge resulting group functions, such as fast turnover of active agents and efficient exchange of information.

\begin{figure*}[tb]
\begin{center}
\includegraphics[width=0.75\textwidth]{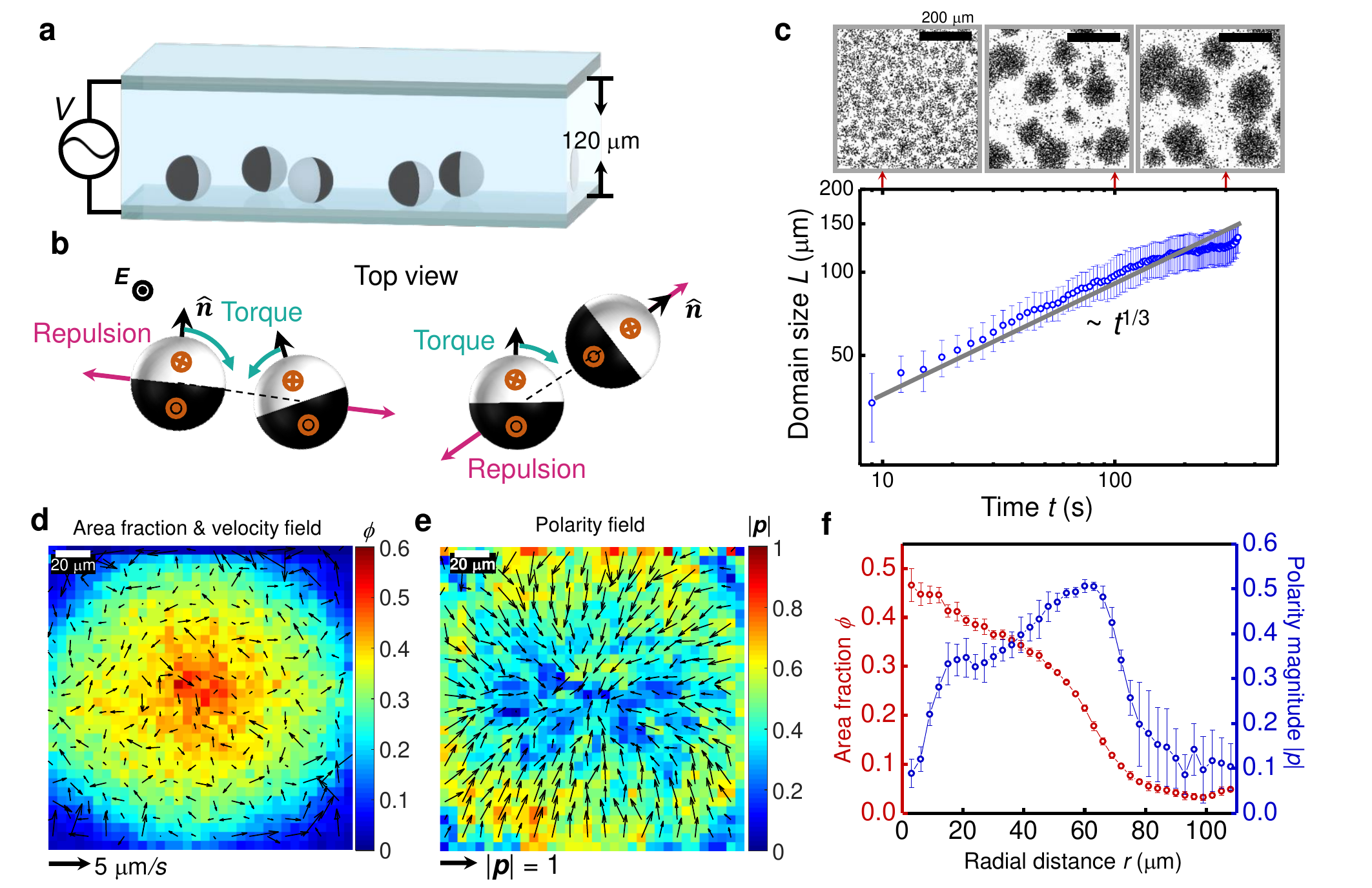}
\end{center}
  {\phantomsubcaption\label{Fig system}}
  {\phantomsubcaption\label{Fig interactions}}
  {\phantomsubcaption\label{Fig coarsening}}
  {\phantomsubcaption\label{Fig density}}
  {\phantomsubcaption\label{Fig polarity}}
  {\phantomsubcaption\label{Fig profiles}}
\bfcaption{Active phase separation in metal-dielectric Janus colloids}{ \subref*{Fig system}, Schematic of the experimental setup in which $3$ $\mu$m-diameter particles are allowed to sediment in water to the bottom of a sample cell across which AC electric fields are applied vertically. \subref*{Fig interactions}, Top view of two Janus particle pairs in an electric field that induces dipoles of opposite orientation and different magnitude (orange) on the head and tail hemispheres. This leads to particle self-propulsion along the direction $\hat{\bm{n}}$ (black), and to interparticle forces (purple) and torques (green). Torques rotate particles in the direction of the interparticle distance, which is indicated by the dashed line. These torques are generally non-reciprocal. \subref*{Fig coarsening}, Clusters coarsen (\hyperref[movies]{Movie 1}, $30$ kHz, $83$ V/mm) with domain-growth kinetics compatible with the Lifshitz-Slyozov relation $L(t) \sim t^{1/3}$. Error bars are S.D. over four independent experiments. \subref*{Fig density},\subref*{Fig polarity}, Time-averaged local area fraction (\subref*{Fig density}, color), along with velocity field (\subref*{Fig density}, arrows) and polarity field (\subref*{Fig polarity}, arrows with magnitude in color) in a cluster ($30$ kHz, $66$ V/mm). Averaging is over $6$ $\mu$m square bins over $50$ s ($2500$ frames), during which the number of particles in the cluster remains approximately constant (\cref{Fig number-fluctuations}). \subref*{Fig profiles}, Angle-averaged radial profiles of area fraction (red) and polarity magnitude (blue) corresponding to panels \subref*{Fig density} and \subref*{Fig polarity}. Error bars are S.D.} \label{Fig 1}
\end{figure*}

\begin{figure}[tb]
\begin{center}
\includegraphics[width=\columnwidth]{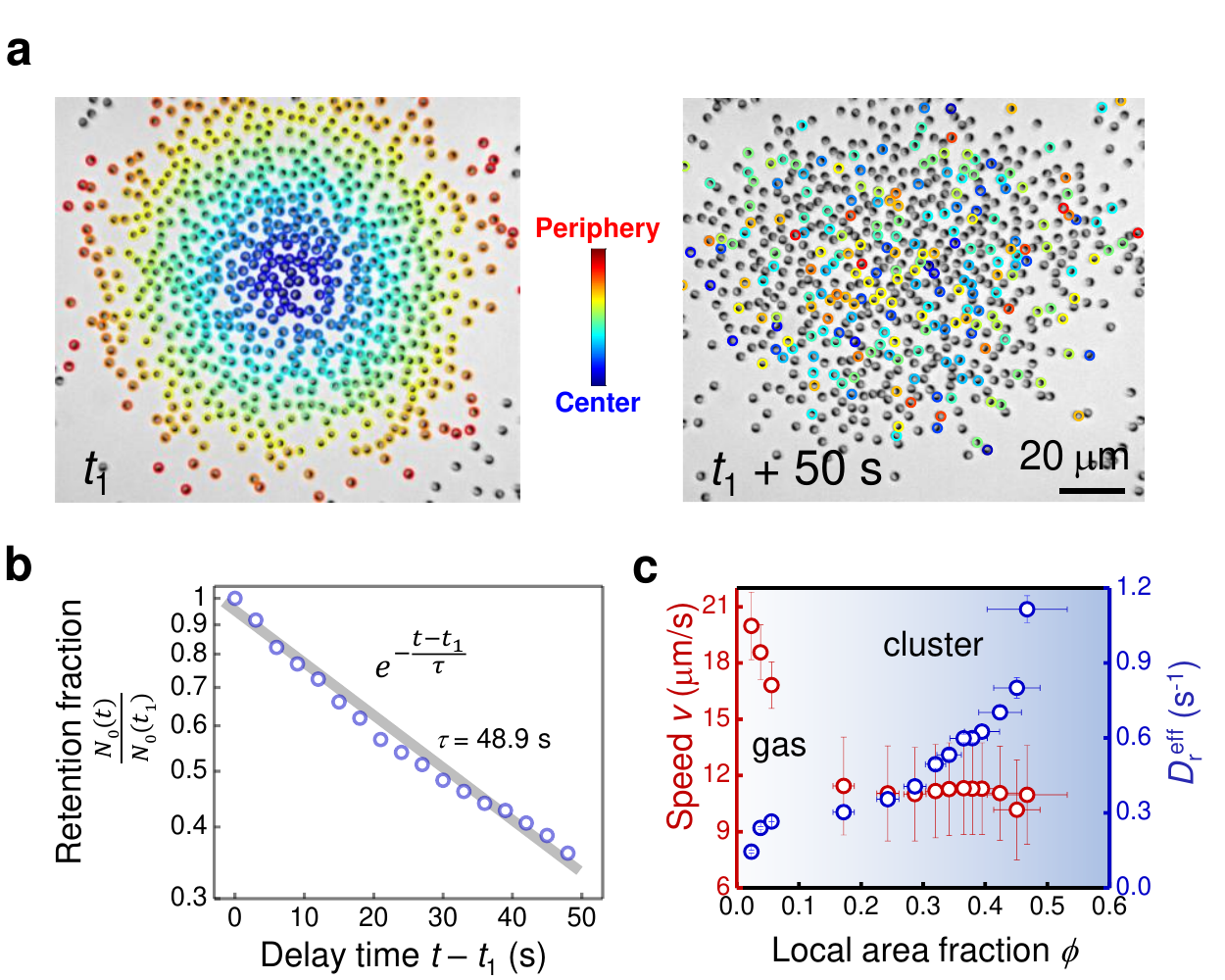}
\end{center}
  {\phantomsubcaption\label{Fig turnover}}
  {\phantomsubcaption\label{Fig retention}}
  {\phantomsubcaption\label{Fig speed-rotation}}
\bfcaption{Particle turnover dynamics of non-jammed clusters whose population remains nearly constant}{ \subref*{Fig turnover}, Example of particle turnover between the gas and cluster states for a cluster whose total population is nearly constant over the experimental time window (\hyperref[movies]{Movie 3}). The first snapshot shows the cluster at an initial time $t=t_1$, chosen arbitrarily, with particles colored according to their distance from the cluster centroid. In the second snapshot, $50$ s later, particles have moved through the cluster and turned over. \subref*{Fig retention}, Particle retention dynamics. The fraction of initial particles that remain within the cluster decays exponentially with elapsed time. \subref*{Fig speed-rotation}, Average particle speed (red) and effective rotational diffusivity (blue) within the gas and cluster states, as functions of the local area fraction. The three points at low area fractions ($\phi < 0.1$) correspond to the gas state. Increasingly higher area fractions correspond to regions deeper into clusters. The speed and effective rotational diffusivity are obtained from the first 0.5 s of the translational and angular mean squared displacements, respectively (\cref{Fig MSD}). Averaging is over $50$ s ($2500$ frames) in the cluster shown in panel \subref*{Fig turnover}. Error bars are S.D.} \label{Fig 2}
\end{figure}

\bigskip
\noindent\textbf{Active phase separation in metal-dielectric Janus colloids}

Here, we combine theory and experiments on self-propelled Janus particles driven by dielectrophoresis. The particles are $3$ $\mu$m-diameter silica spheres, coated with titanium (\hyperref[methods]{Methods}) on one hemisphere. These particles are suspended in a $0.05$ mM NaCl aqueous solution and placed between conductive coverslips coated with indium tin oxide, separated by a $120$ $\mu$m spacer (\cref{Fig system}, \hyperref[methods]{Methods}). Particles sediment to form a dilute monolayer with area fraction in the range $\phi_0 \approx 0.05-0.15$. To drive the particles, we apply a perpendicular AC voltage of amplitude $V_0=8-10$ V and frequency $\nu=30$ kHz. The resulting electric field tends to align the particle equator perpendicular to the coverslips. The resulting unequal electric polarization on the metal and dielectric hemispheres (\cref{Fig interactions}) induces electrokinetic flows that produce particle self-propulsion \cite{Gangwal2008,Moran2017,Yan2016} (along a direction $\hat{\bm{n}}$ pointing from the metallic to the dielectric hemisphere), as well as electrostatic interparticle forces and torques (\cref{Fig interactions}).

Clusters, observed within seconds after switching on the electric field, coarsen in a process suggestive of Ostwald ripening, with large clusters growing and small clusters shrinking and disappearing (\hyperref[movies]{Movie 1}, \cref{Fig ripening}). Domain-growth kinetics are compatible with the Lifshitz-Slyozov relation $L(t) \sim t^{1/3}$ of classic phase separation (\cref{Fig coarsening}), consistent with the mapping to an effective free energy we present below.

As some particles stuck on the coverslip were not possible to avoid experimentally, it was natural to inquire whether clusters necessarily nucleated around them. This possibility was discounted as at early times the majority of clusters encompassed no stuck particles (\cref{Fig stuck-particles}). On the other hand, during coarsening a large fraction of clusters contain a few particles that are stuck on the coverslip, which might help form and stabilize the clusters (\cref{Fig stuck-particles}).

\bigskip
\noindent\textbf{Clusters have wide interfaces}

Despite the familiar coarsening kinetics, the structure and dynamics of individual clusters differ markedly from those observed in passive phase separation and repulsion-based MIPS. Rather than displaying the standard uniform bulk and sharp interface, our clusters exhibit a pronounced density gradient (\cref{Fig density,Fig cluster-growth}, and \hyperref[movies]{Movie 2}) and inward-pointing polarity (\cref{Fig polarity}), defined as $\bm{p} = \langle \hat{\bm{n}}\rangle$. Particle density increases from the edge to the center without a density plateau (\cref{Fig profiles}, red) for clusters up to $\sim 300$ $\mu$m in diameter, suggesting very wide interfaces, at least several tens of micrometers. The central density (area fraction $\phi\sim 0.6$) never approaches close packing. Reciprocally, polarity $|\bm{p}|$ is highest at the cluster edge and decreases toward the center (\cref{Fig profiles}, blue). This decrease suggests that clusters might eventually grow large enough to develop an isotropic ($\bm{p} = 0$) bulk phase. The inward-pointing polarity at clusters’ edges prevents them from coalescing immediately upon contact. Rather, upon collisions between clusters, visible boundaries persist for up to minutes (\hyperref[movies]{Movie 1}, see snapshots in \cref{Fig coarsening}).

\bigskip
\noindent\textbf{Non-jammed clusters with fast particle turnover}

The dynamics of individual particles within clusters differs markedly from that in jamming-based MIPS: Clusters are fluid, not jammed. Particles move easily through clusters (\hyperref[movies]{Movie 3}), which exhibit fast turnover, with particles leaving and joining a cluster on a time scale of tens of seconds (\cref{Fig turnover}). The number of native particles that remain in a cluster decays exponentially with a characteristic time depending on the cluster size and particle speed (\cref{Fig retention,Fig turnover-size}, and \hyperref[movies]{Movie 4}). By tracking particle position and orientation at different local area fractions $\phi$, we obtain their mean square translational and angular displacements (\cref{Fig MSD}). From these measurements, we find that particles within clusters ($\phi >0.1$) are slower than those outside them ($\phi < 0.1$), but do not slow down further at the higher densities deeper inside clusters (\cref{Fig speed-rotation}, red). On the other hand, the effective rotational diffusivity $D_{\text{r}}^{\text{eff}}$ increases monotonically with local area fraction (\cref{Fig speed-rotation}, blue), indicating faster particle reorientations due to stronger interparticle torques in denser regions.

\begin{figure*}[tb]
\begin{center}
\includegraphics[width=0.75\textwidth]{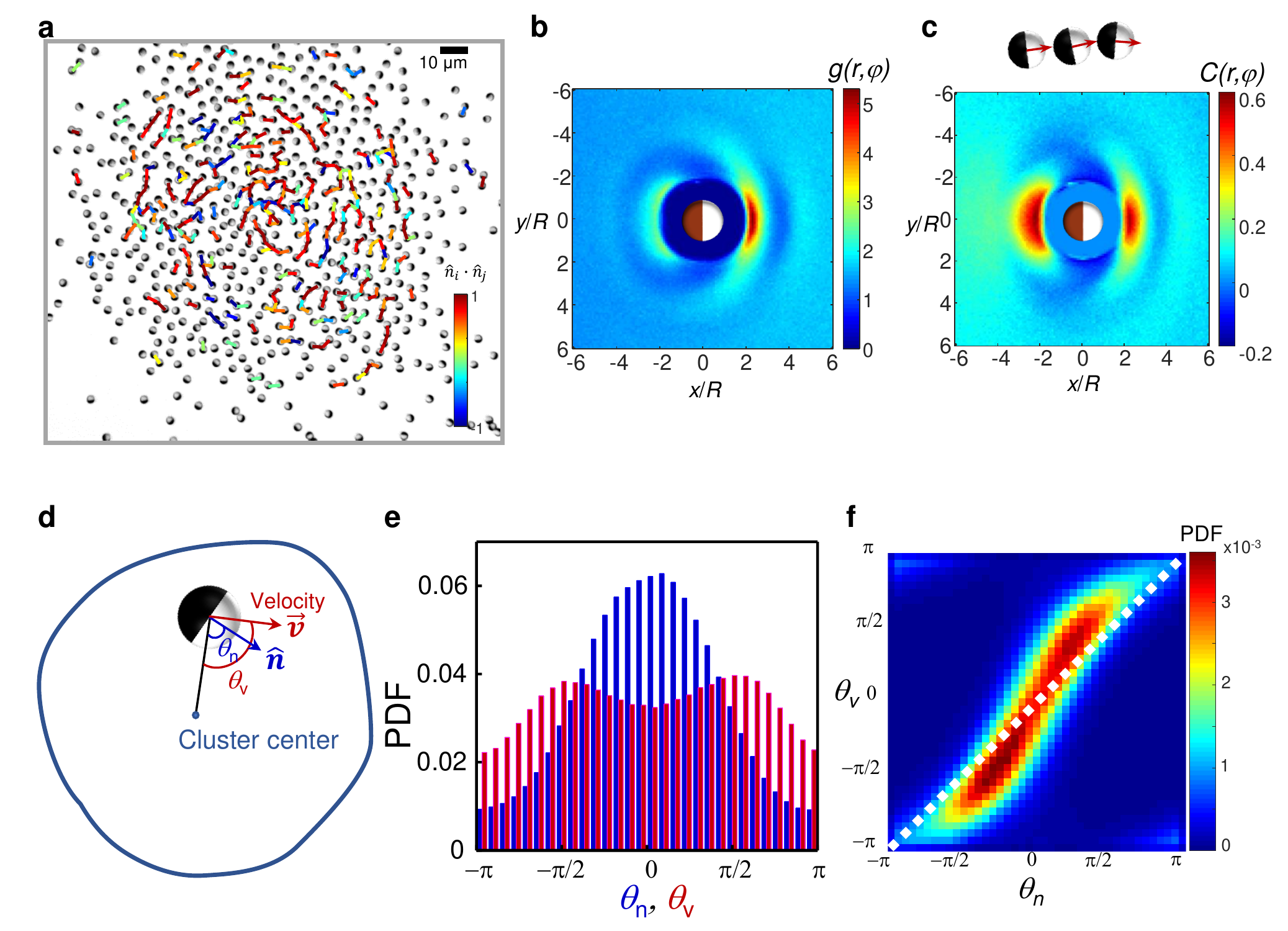}
\end{center}
  {\phantomsubcaption\label{Fig orientations}}
  {\phantomsubcaption\label{Fig distribution-function}}
  {\phantomsubcaption\label{Fig orientation-correlations}}
  {\phantomsubcaption\label{Fig angles-definitions}}
  {\phantomsubcaption\label{Fig angles-histograms}}
  {\phantomsubcaption\label{Fig angles-joint-histogram}}
\bfcaption{Particle orientation, velocity, and correlations in non-jammed clusters}{ \subref*{Fig orientations}, Snapshot of a cluster, overlaid with lines connecting pairs of particles whose centers are separated by less than 3 particle radii. The color map shows the degree of pair alignment, with red indicating chaining. \subref*{Fig distribution-function},\subref*{Fig orientation-correlations}, Pair distribution $g(\bm{r})$ and orientation correlation $C(\bm{r})$ averaged over the cluster for $50$ s ($2500$ frames), showing the tendency of particles to line up and align in chains, respectively. Coordinates are scaled by the particle radius. \subref*{Fig angles-definitions}, Definitions of the angle formed by particle orientation ($\theta_{\text{n}}$) and velocity ($\theta_{\text{v}}$) with respect to the radial direction. Schematically, the blue contour indicates the cluster periphery. \subref*{Fig angles-histograms},\subref*{Fig angles-joint-histogram}, Individual (\subref*{Fig angles-histograms}) and joint (\subref*{Fig angles-joint-histogram}) probability distribution functions (PDFs) of the orientation and velocity angles defined in \subref*{Fig angles-definitions}. In \subref*{Fig angles-joint-histogram}, deviation from the diagonal shows misalignment between particle orientation and velocity.} \label{Fig 3}
\end{figure*}

\bigskip
\noindent\textbf{Flickering chains facilitate particle motion}

Since head and tail particle hemispheres attract each other (\cref{Fig interactions}), particles in clusters often form chains, 3-7 particles long, which last hundreds of milliseconds (\hyperref[movies]{Movie 5}, red lines in \cref{Fig orientations}). These chains constantly deform, break, and reform, with particles hopping on and off different chains. To characterize positional and orientational order, we measure the pair distribution function $g(\bm{r}) = \phi(\bm{r})/\phi_0$ and the orientation correlation function $C(\bm{r}) = \langle \hat{\bm{n}}(\bm{r}') \cdot \hat{\bm{n}}(\bm{r}'+\bm{r})\rangle$. As expected for self-propelled particles, $g(\bm{r})$ is anisotropic; it is more likely to find another particle ahead than behind a reference particle (\cref{Fig distribution-function}). We also find that it is more likely to find another particle behind than on the side of the reference particle, producing a depletion wing pattern (\cref{Fig orientation-correlations}). While recent work showed that these depletion wings can arise even in the absence of alignment interactions \cite{Poncet2021}, in our system they result from torques generated by head-tail attraction. Finally, orientational correlations are stronger along the direction of self-propulsion than perpendicular to it (\cref{Fig orientation-correlations}), showing that particles tend to align along the chains, but not with lateral neighbors.

\bigskip
\noindent\textbf{Particle orientation and velocity are misaligned in clusters}

Despite these transient chains, particle motion is disordered at long times, as shown by the time-averaged velocity field (\cref{Fig density}, arrows). Interestingly, this lack of velocity order coexists with radial polar order (\cref{Fig polarity}). This distinction is apparent in the probability distributions of the angles formed by the particle orientation and velocity with respect to the clusters’ radial direction (\cref{Fig angles-definitions}): Whereas the orientation angle distribution peaks at $0$ (\cref{Fig angles-histograms}, blue), consistent with radial order, the velocity angle distribution peaks at $\pm \pi/2$ (\cref{Fig angles-histograms}, red), indicating flows orthogonal to the radial direction. I.e., even though particles orient mainly towards the center of the cluster, they have a higher chance to move tangentially to it. As expected, the difference between orientation and velocity angles is absent prior to cluster formation (\cref{Fig angles-joint-histogram-uniform}), but emerges from the stronger interparticle interactions within clusters (\cref{Fig angles-joint-histogram}).

\begin{figure*}[tb]
\begin{center}
\includegraphics[width=0.75\textwidth]{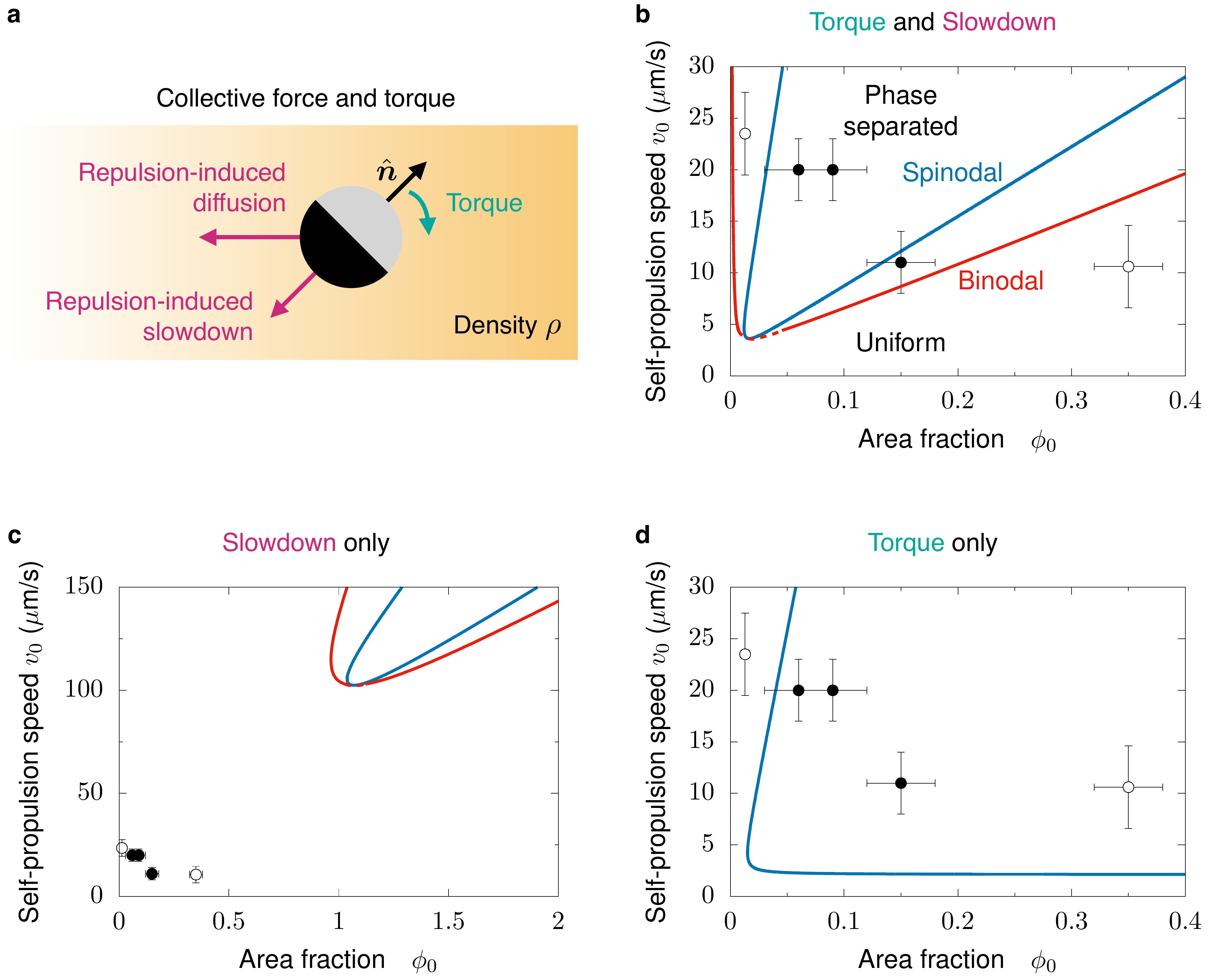}
\end{center}
  {\phantomsubcaption\label{Fig collective}}
  {\phantomsubcaption\label{Fig diagram}}
  {\phantomsubcaption\label{Fig slowdown-only}}
  {\phantomsubcaption\label{Fig torques-only}}
\bfcaption{State diagram of torque-based active phase separation}{ \subref*{Fig collective}, In the theoretical model, a probe particle experiences collective interaction force and torque (\cref{eq collective-main}) in a particle density field $\rho(\bm{r})$, which increases from left to right in this depiction. Repulsion leads to two collective forces (purple): one opposite to self-propulsion (slowdown), and one opposite to the density gradient (diffusion). Collective torques (green) tend to align particle motion with density gradients. \subref*{Fig diagram}, State boundaries (curves) predicted using experimental estimates for parameter values (\cref{t parameters} in \cref{parameters}). The numerically-unresolved region of the binodal near the critical point is indicated by a dashed curve as a guide to the eye. The filled and open data points show the experimental conditions for which phase separation was observed or not, respectively. \subref*{Fig slowdown-only}, State diagram predicted from slowdown only, i.e. $\tau_1 = 0$. Note the different scale from panel \subref*{Fig diagram}. \subref*{Fig torques-only}, State diagram predicted from torques only, i.e. $\zeta_0 = 0$. Note the absence of binodal in this case. Error bars are S.D. obtained as in \cref{Fig speed-rotation}.} \label{Fig 4}
\end{figure*}

\bigskip
\noindent\textbf{Model for Janus particles with electrostatic interactions}

As our experiments show MIPS without substantial self-trapping, non-standard mechanisms might be responsible for phase separation in our system. We therefore developed a microscopic model based on the dipolar interactions between the hemispheres of our particles (\cref{microscopic-model,Fig Janus}). Our model shows that two particles interact via a repulsive force
\begin{equation} \label{eq force}
\bm{F}_{ij} \approx \frac{3}{4\pi\epsilon} \frac{(d_{\text{h}} + d_{\text{t}})^2}{r_{ij}^4} e^{-r_{ij}/\lambda}\; \hat{\bm{r}}_{ij},
\end{equation}
where $\epsilon$ is the dielectric permittivity of the solvent, $\bm{r}_{ij} = \bm{r}_j - \bm{r}_i$ is the distance vector, and $d_{\text{h}}<0$ and $d_{\text{t}}>0$ are the effective dipole strengths of the head and tail hemispheres, respectively (\cref{Fig interactions}). The exponential factor accounts for screening by the electrodes, separated by a distance $\lambda = 120$ $\mu$m. Moreover, because tail dipoles are stronger than head dipoles ($d_{\text{t}}^2> d_{\text{h}}^2$), particles interact via a torque
\begin{equation} \label{eq torque}
\bm{\Gamma}_{ij} \approx \frac{3\ell}{4\pi\epsilon} \frac{d_{\text{h}}^2 - d_{\text{t}}^2}{r_{ij}^4} e^{-r_{ij}/\lambda}\; \hat{\bm{n}}_j\times \hat{\bm{r}}_{ij},
\end{equation}
where $\ell = 3R/8$ is the distance by which the dipoles are off-centered, with $R=1.5$ $\mu$m the particle radius. This torque tends to reorient the particles in the direction of the interparticle distance vector $\bm{r}_{ij}$ (\cref{Fig interactions}). Hence, this torque is responsible for chain formation (\cref{Fig 3}); particles aligned in a chain experience no torque because they point along $\bm{r}_{ij}$ (\cref{Fig interactions}, right). Even though the underlying forces between dipoles are reciprocal, the torques $\bm{\Gamma}_{ij}$ between particles are in general non-reciprocal: $\bm{\Gamma}_{ij}\neq -\bm{\Gamma}_{ji}$. Whereas one particle may be already aligned with the interparticle distance $\bm{r}_{ij}$, the other one may not (\cref{Fig interactions}, right). As two particles reorient in a non-reciprocal way, they also rotate around their common center of mass (\cref{microscopic-model}). Experimentally, torque non-reciprocity manifests in the dynamics and statistics of two-particle interaction events (\cref{Fig interaction-statistics}).

We write Langevin equations for the translational and rotational motion of particle $i$ as
\begin{subequations} \label{eq Langevin-main}
\begin{align}
\frac{\dd\bm{r}_i}{\dd t} &= v_0 \hat{\bm{n}}_i + \frac{\bm{F}_i}{\xi_{\text{t}}} + \bm{\eta}^{\text{t}}_i(t);\qquad \bm{F}_i = \sum_{j\neq i} \bm{F}_{ji}, \label{eq Langevin-translation-main}\\
\frac{\dd\hat{\bm{n}}_i}{\dd t} &= \frac{\bm{\Gamma}_i}{\xi_{\text{r}}} + \bm{\eta}_i^{\text{r}}(t);\qquad \bm{\Gamma}_i= \sum_{j\neq i} \bm{\Gamma}_{ji},\label{eq Langevin-rotation-main}
\end{align}
\end{subequations}
where $v_0$ is the self-propulsion speed, $\xi_{\text{t}}$ and $\xi_{\text{r}}$ are the translational and rotational friction coefficients, respectively, and $\bm{\eta}^{\text{t}}_i(t)$ and $\bm{\eta}^{\text{r}}_i(t)$ are both Gaussian white noise (\cref{microscopic-model}).

\bigskip
\noindent\textbf{Collective forces and torques}

To predict the collective behavior of the system, we coarse-grain the microscopic model (\cref{coarse-graining}) and obtain the Smoluchowski equation for the probability $\Psi_1 (\bm{r},\hat{\bm{n}};t)$ of finding a particle at position $\bm{r}$ and orientation $\hat{\bm{n}}$ at time $t$ (\cref{eq psi1}). This probability evolves under the action of collective interaction forces and torques that depend on the particle density field $\rho(\bm{r})$ \cite{Speck2020,Kirkwood1949,Irving1950}. To first order in density gradients, we obtain (\cref{coarse-graining})
\begin{subequations} \label{eq collective-main}
\begin{align}
\bm{F}_{\text{int}}(\bm{r},\hat{\bm{n}}) &= -\Psi_1(\bm{r},\hat{\bm{n}}) \left[\zeta_0 \rho(\bm{r})\hat{\bm{n}} + \zeta_1 \bm{\nabla}\rho(\bm{r})\right],\label{eq collective-force-main}\\
\bm{\Gamma}_{\text{int}}(\bm{r},\hat{\bm{n}}) &= \Psi_1(\bm{r},\hat{\bm{n}})\, \tau_1\, \hat{\bm{n}}\times \bm{\nabla}\rho(\bm{r}) \label{eq collective-torque-main}.
\end{align}
\end{subequations}
The coefficients $\zeta_0,\zeta_1,\tau_1>0$ (\cref{eq coefficients}) depend on the pair distribution function $g(\bm{r})$ in the uniform state, which we measure in experiments (\cref{Fig gr}). Due to the higher probability of finding other particles in front of rather than behind the probe particle, the first contribution in \cref{eq collective-force-main} gives a repulsion-induced force that opposes self-propulsion \cite{Bialke2013,Speck2020} (\cref{Fig collective}). The higher the particle density, the higher the opposing force; the resulting density-induced slowdown produces standard self-trapping MIPS familiar from extensive theoretical study \cite{Cates2015,Gonnella2015,Marchetti2016,Zottl2016,Speck2020,Ma2020,Fily2012,Redner2013,Bialke2013}. The second contribution in \cref{eq collective-force-main} predicts a repulsion-induced force against density gradients, tending to homogenize particle concentration like a diffusive flux (\cref{Fig collective}). Finally, the collective torque in \cref{eq collective-torque-main} tends to align particle orientation $\hat{\bm{n}}$ with the density gradient, thus reorienting particle motion \emph{toward} higher-density regions (\cref{Fig collective}), as observed at the clusters edges in experiments (\cref{Fig density,Fig polarity,Fig profiles}). This collective torque requires non-reciprocity of the interparticle torques (\cref{eq torque}). Reciprocal torques, such as $\bm{\Gamma}_{ij} \propto \hat{\bm{n}}_i \times \hat{\bm{n}}_j$, could not orient particles toward the location of other particles, and hence $\bm{\Gamma}_{\text{int}}$ would vanish \cite{Jayaram2020,Grossmann2020} (\cref{coarse-graining}). Finally, while non-reciprocal torques can lead to chiral phases \cite{Fruchart2021}, we do not find them here.

\bigskip
\noindent\textbf{Torque-based phase separation}

As particles reorient toward crowded areas, they self-propel up their own density gradient (\cref{Fig collective}). Hence, particles migrate toward crowded regions, which produces an instability promoting phase separation. Similar behavior was observed in active agents with finite vision cones \cite{Barberis2016,Durve2018,Lavergne2019}. To predict the instability, we complete the coarse-graining and obtain hydrodynamic equations (\cref{coarse-graining}). The density field follows a continuity equation,
\begin{equation} \label{eq density-main}
\partial_t \rho = -\bm{\nabla}\cdot \bm{J};\qquad \bm{J} = v[\rho] \bm{p} - (D_{\text{t}} + D_{\text{rep}}[\rho])\bm{\nabla}\rho,
\end{equation}
where the flux includes contributions from self-propulsion at a density-dependent speed $v[\rho(\bm{r})] = v_0 - \zeta_0 \rho(\bm{r}) /\xi_{\text{t}}$, and diffusion that combines both bare and repulsion-induced diffusivities, $D_{\text{t}}$ and $D_{\text{rep}} [\rho(\bm{r})] = \zeta_1 \rho(\bm{r})/\xi_{\text{t}}$, respectively. At times $t\gg D_{\text{r}}^{-1}$, the polarity field $\bm{p}$ becomes slaved to the density field (\cref{phase-diagram}):
\begin{equation} \label{eq adiabatic-polarity-main}
\bm{p} = \frac{1}{2D_{\text{r}}} \left( v_{\text{tor}} [\rho] \bm{\nabla}\rho - \bm{\nabla}(v[\rho] \rho)\right),
\end{equation}
where the density-dependent speed $v_{\text{tor}} [\rho(\bm{r})] = \tau_1 \rho(\bm{r})/\xi_{\text{r}}$ embodies the effects of torques in polarizing the system toward increasing densities. Introducing \cref{eq adiabatic-polarity-main} into \cref{eq density-main}, we obtain $\bm{J} = - \mathcal{D}[\rho]\bm{\nabla}\rho$, where
\begin{equation} \label{eq collective-diffusion-main}
\mathcal{D}[\rho] = D_{\text{t}} + D_{\text{rep}}[\rho] + \frac{v[\rho]}{2D_{\text{r}}} \left( v[\rho] + v'[\rho]\rho - v_{\text{tor}}[\rho] \right)
\end{equation}
is a collective diffusivity. Thus, a uniform state with density $\rho_0$ experiences a spinodal instability for $\mathcal{D}(\rho_0) < 0$. In the absence of interaction torques ($v_{\text{tor}}[\rho]=0$), $\mathcal{D}$ can turn negative due to repulsion-induced slowdown ($v'[\rho]<0$), which is the standard mechanism for MIPS \cite{Cates2015}. Here, \cref{eq collective-diffusion-main} shows that, even in the absence of slowdown ($v'[\rho] = 0$), torques alone ($v_{\text{tor}}[\rho]>0$) can produce a MIPS-like instability.

Furthermore, we establish that the torque-induced instability leads to phase coexistence. To this end, we express the particle flux $\bm{J}$ as deriving from an effective chemical potential $\mu[\rho]$ \cite{Cates2015,Stenhammar2013,Wittkowski2014,Speck2014,Speck2015,Paliwal2018,Solon2018,Solon2018b}: $\bm{J} = -M[\rho] \bm{\nabla}\mu[\rho]$, with $M[\rho]$ the mobility functional (\cref{phase-diagram}). We then use the relation $f'(\rho) = \mu(\rho)$ to obtain a local effective free energy $f(\rho)$, which has the conventional double-well shape (\cref{Fig thermodynamics}). Ignoring non-local corrections \cite{Cates2015,Wittkowski2014,Solon2018,Solon2018b}, we use the common-tangent construction on $f(\rho)$ to predict the densities of the coexisting phases, i.e. the binodal lines of the phase diagram (\cref{phase-diagram}). Importantly, our theory predicts that these uniform-density phases have no orientational order (see \cref{eq adiabatic-polarity-main}). Our theory is approximate; hence, we do not expect the predicted binodal lines to be quantitatively accurate. Yet, the existence of phase coexistence is a robust prediction which relies only on two ingredients: the torques toward dense regions and the decrease of particle speed at high densities (\cref{phase-diagram}). In our experiments, clusters do not achieve uniform bulk density (\cref{Fig profiles}), and therefore we are unable to observe the predicted phase coexistence. This fact suggests to us that the experimental system is in a dynamical regime whose asymptotic behavior at very large cluster size was not yet achieved. 

To compare our predictions to experiments, we estimate the values of all the model parameters, including the particles’ translational and rotational diffusion and friction coefficients, as well as the strength of the electric dipoles (\cref{parameters}). These estimates allow us to predict the phase diagram in the conditions of our experiments. Our experimental observations of phase separation fall within the predicted region of the phase diagram when we include both the slowdown and torque effects (\cref{Fig diagram}). Similarly, the absence of phase separation at high area fraction also agrees with our predictions (\cref{Fig diagram}). In the uniform high-density state, we observe transient particle chains throughout the system (\hyperref[movies]{Movie 6}, \cref{Fig high-density-uniform}). With slowdown only, we cannot account for our experimental observations; the predicted phase-separation region lies at much higher self-propulsion speeds and densities than experimentally observed (\cref{Fig slowdown-only}). Conversely, while torques alone can account for the instability of the uniform phase (spinodal in \cref{Fig torques-only}), they do not yield phase coexistence (no binodal in \cref{Fig torques-only}). Repulsion-induced slowdown is required to stabilize the dense phase.

In summary, we have demonstrated a new type of active phase separation based on non-reciprocal torques. Active agents reorient themselves toward crowded areas to form structured clusters, while moving easily within clusters and also into and out of them. Perhaps more fundamentally, our theory shows that orientational interactions (torques) can produce phases of matter without internal orientational order. Our work thus establishes connections between the paradigms of aligning and non-aligning active matter, contributing to the understanding of how different types of interparticle interactions can yield qualitatively new kinds of collective nonequilibrium phenomena \cite{Marchetti2013,Hagan2016,Fodor2018,Shaebani2020}.


\bigskip

\noindent\textbf{Acknowledgments}

J.Z. and S.G. were supported by the taxpayers of South Korea through the Institute of Basic Science, project code IBS-R020-D1. R.A. thanks Julien Tailleur for insightful discussions, and acknowledges discussions with the participants of the virtual ``Active 20'' KITP program, supported in part by the National Science Foundation under Grant No. NSF PHY-1748958. R.A. acknowledges support from the Human Frontier Science Program (LT000475/2018-C). J.Y. holds a Career Award at the Scientific Interface from the Burroughs Wellcome Fund. N.S.W. acknowledges support from the National Science Foundation, through the Center for the Physics of Biological Function (PHY-1734030).

\bigskip

\noindent\textbf{Author contributions}

J.Z., J.Y., and S.G. conceived the experiment. J.Z. performed the experiments and analyzed data with help from J.Y. R.A. conceived and developed the theory and analyzed data. N.S.W. supervised the theory. All authors discussed and interpreted the results. J.Z., R.A., N.S.W., and S.G. wrote the manuscript.

\bigskip

\noindent\textbf{Competing interests}

The authors declare no competing interests.

\bigskip

\noindent\textbf{Data availability}

All data are available from the authors upon request.

\bigskip

\noindent\textbf{Code availability}

All codes are available from the authors upon request.

\bibliography{Torques}

\onecolumngrid 

\clearpage

\twocolumngrid

\section*{Methods} \label{methods}

\subsection*{Particle synthesis} \label{synthesis}

Following protocols described elsewhere \cite{Yan2016}, a submonolayer of $3$ $\mu$m-diameter silica particles (Tokuyama) is prepared on a standard glass slide. To obtain metal-dielectric Janus particles, $20$ nm of titanium and then $5$ nm of SiO$_2$ are deposited vertically on the glass slide using an electron-beam evaporator. The preparation is then washed with isopropyl alcohol and deionized water, and then sonicated into deionized water to collect the Janus particles.

\subsection*{Experimental setup} \label{setup}

NaCl stock solution is added to the particle suspension to obtain $0.05$ mM NaCl solutions. The particle suspensions are confined between two coverslips (SPI Supplies) coated with indium tin oxide to make them conductive, and with $25$ nm of silicon oxide to prevent particles from sticking to them. The coverslips have a $9$ mm hole in the center, separated by a $120$ $\mu$m-thick spacer (GraceBio SecureSeal). An alternating voltage is applied between the coverslips using a function generator (Agilent 33522A). The sample cell is imaged with 5X and 40X air objectives on an inverted microscope (Axiovert 200). Microscopic images and videos are taken with a CMOS camera (Edmund Optics 5012M GigE) with $20$ ms time resolution.

\subsection*{Image analysis} \label{image-analysis}

Image processing is performed using MATLAB with home-developed codes, which are available upon request to the corresponding author.

\clearpage

\setcounter{equation}{0}
\setcounter{figure}{0}
\renewcommand{\theequation}{S\arabic{equation}}
\renewcommand{\thefigure}{S\arabic{figure}}

\onecolumngrid
\begin{center}
\textbf{\large Supplementary Information}
\end{center}

\section*{Supplementary Figures} \label{SI figures}

\begin{figure*}[!htb]
\begin{center}
\includegraphics[width=0.9\textwidth]{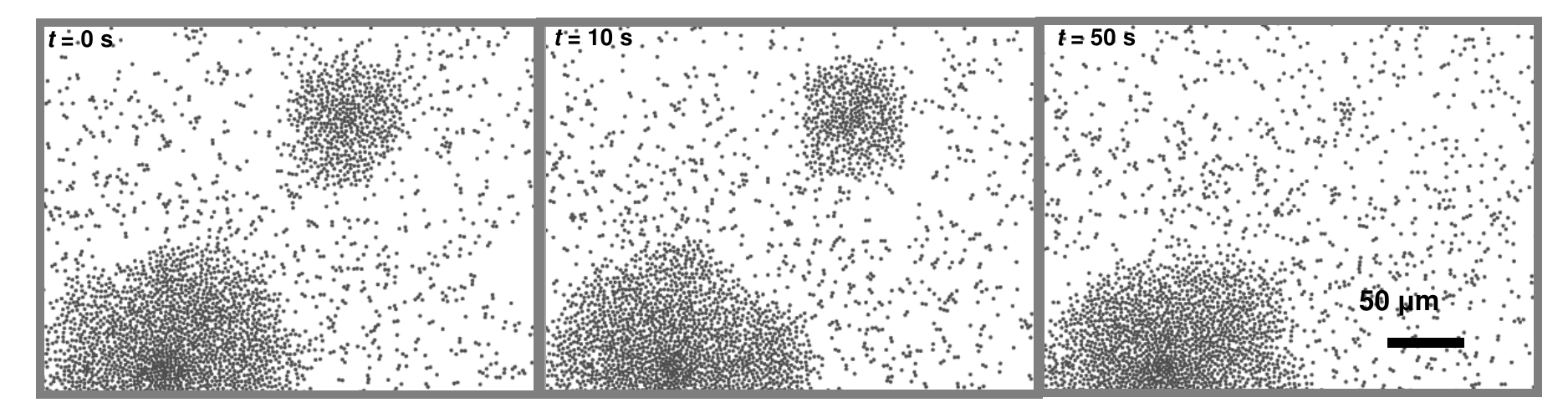}
\end{center}
\bfcaption{Ostwald ripening of active colloids}{ Series of snapshots showing a small cluster shrinking and disappearing as part of the coarsening process.} \label{Fig ripening}
\end{figure*}

\begin{figure*}[!htb]
\begin{center}
\includegraphics[width=0.75\textwidth]{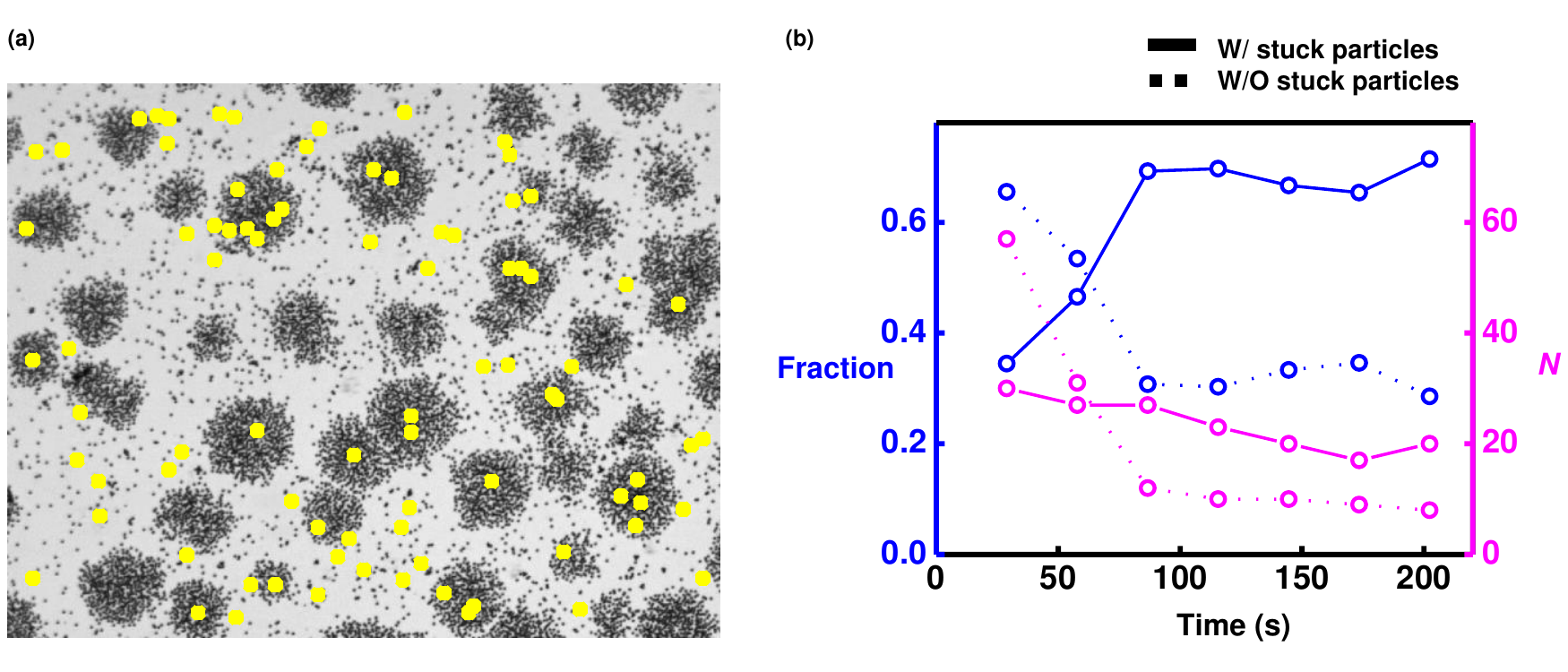}
\end{center}
  {\phantomsubcaption\label{Fig stuck-image}}
  {\phantomsubcaption\label{Fig stuck-fraction}}
\bfcaption{Clusters with and without stuck particles}{ \subref*{Fig stuck-image}, A bright field microscopy image of active phase separation in a late stage with yellow stars labelling particles that, stuck to the coverslip, remain in the same location throughout the experiment. \subref*{Fig stuck-fraction}, The number (magenta) and fraction (blue) of clusters with and without (solid and dotted curves, respectively) at least one stuck particle. In the beginning of the experiment, more clusters are formed without than with stuck particles. With elapsed time, the number of clusters forming both with and without stuck particles decreases (magenta), but clusters with stuck particles become the most abundant.} \label{Fig stuck-particles}
\end{figure*}

\begin{figure*}[!htb]
\begin{center}
\includegraphics[width=0.75\textwidth]{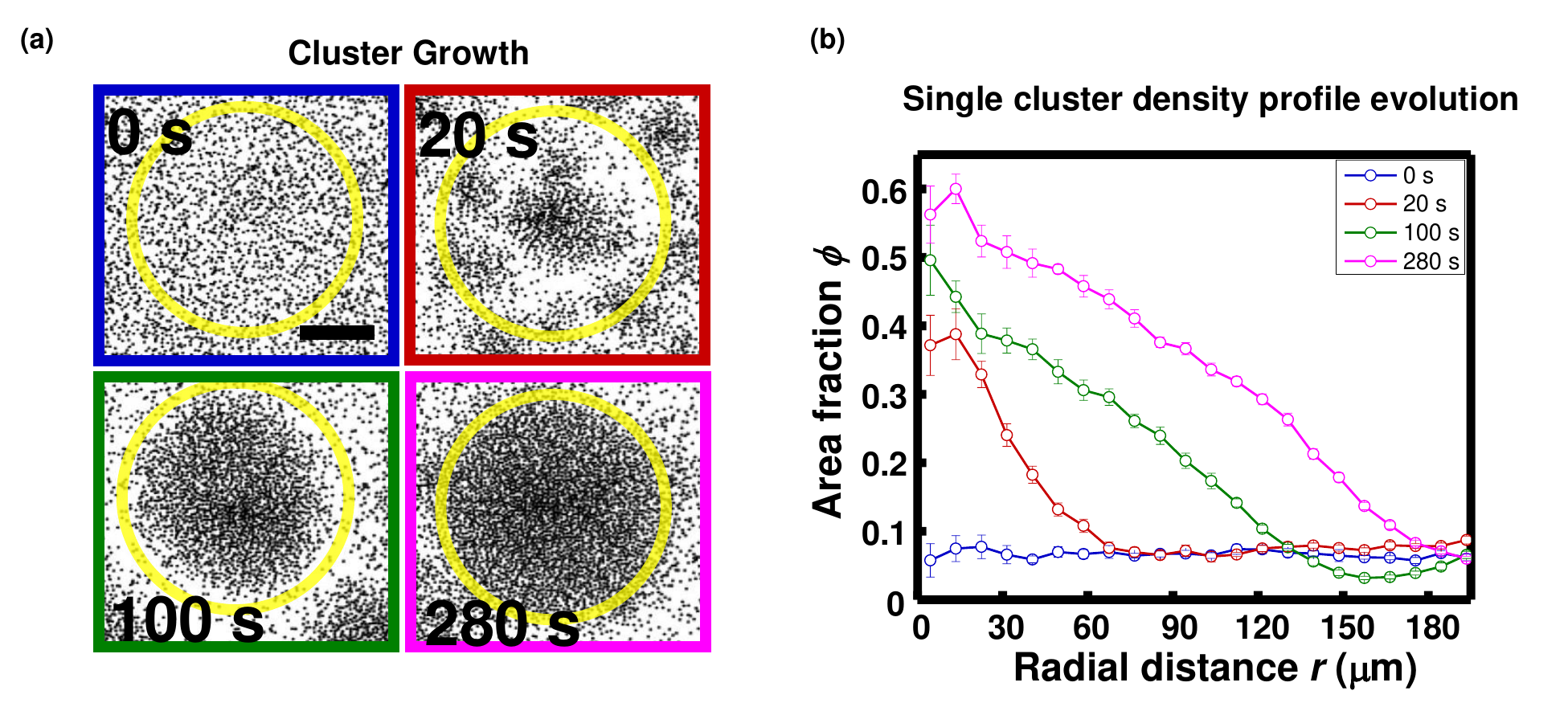}
\end{center}
  {\phantomsubcaption\label{Fig growth-snapshots}}
  {\phantomsubcaption\label{Fig growth-profiles}}
\bfcaption{Dynamics of cluster growth}{ \subref*{Fig growth-snapshots}, Snapshots of cluster growth upon AC electric field application (\hyperref[movies]{Movie 2}). Scale bar, $100$ $\mu$m. The yellow circumference indicates the outline of the cluster in the final snapshot. \subref*{Fig growth-profiles}, Density profile evolution in the growing cluster shown in \subref*{Fig growth-snapshots}. Averages are over $20$ frames ($1.2$ s) centered at each designated time point. Error bars are S.D.} \label{Fig cluster-growth}
\end{figure*}

\begin{figure*}[!htb]
\begin{center}
\includegraphics[width=0.4\textwidth]{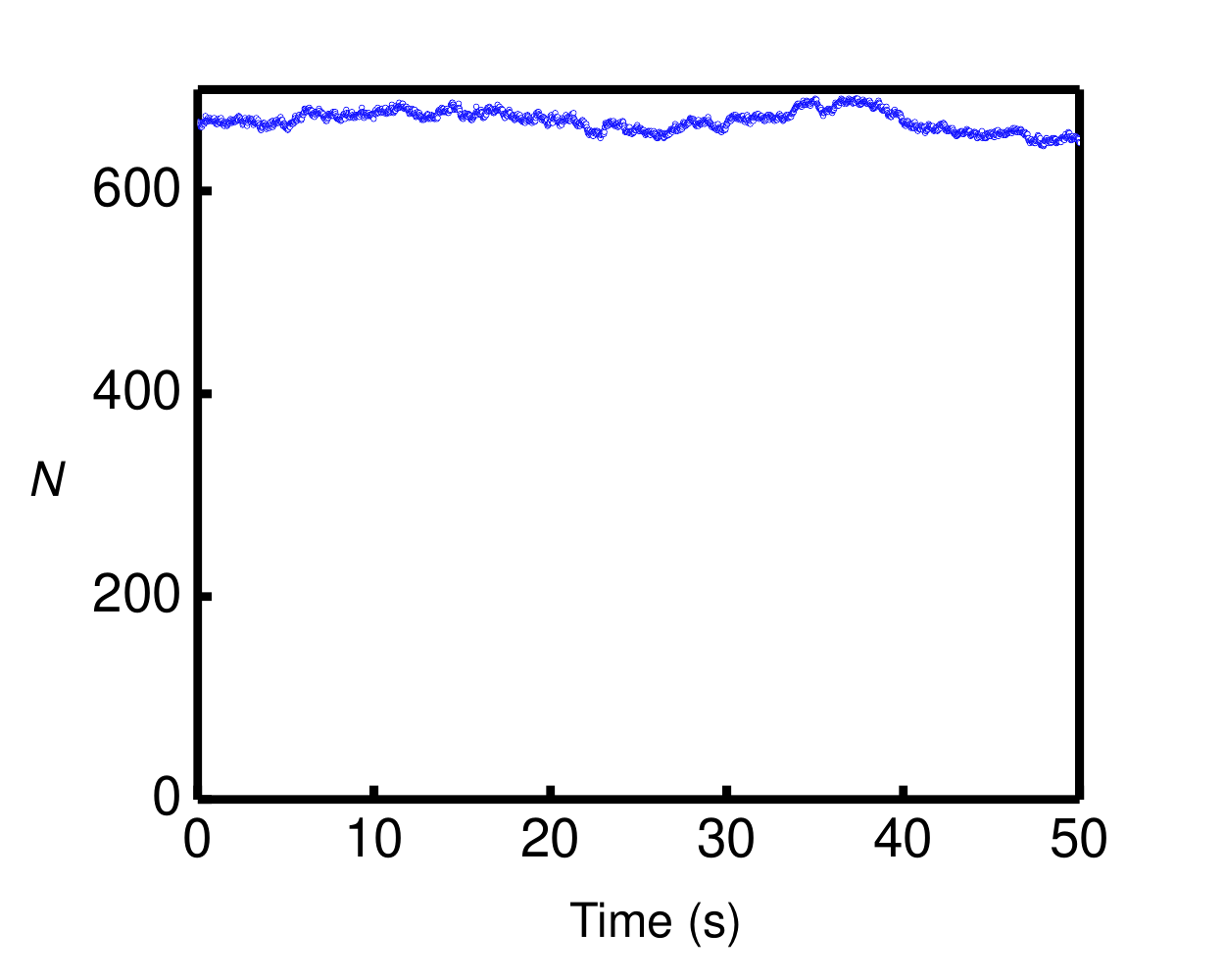}
\end{center}
\bfcaption{Particle-number fluctuations in a quasi-steady state cluster}{ The number of particles in this cluster, the same cluster also shown in \cref{Fig 1,Fig 2}, remains roughly constant during the time when averages are taken.} \label{Fig number-fluctuations}
\end{figure*}

\begin{figure*}[!htb]
\begin{center}
\includegraphics[width=0.4\textwidth]{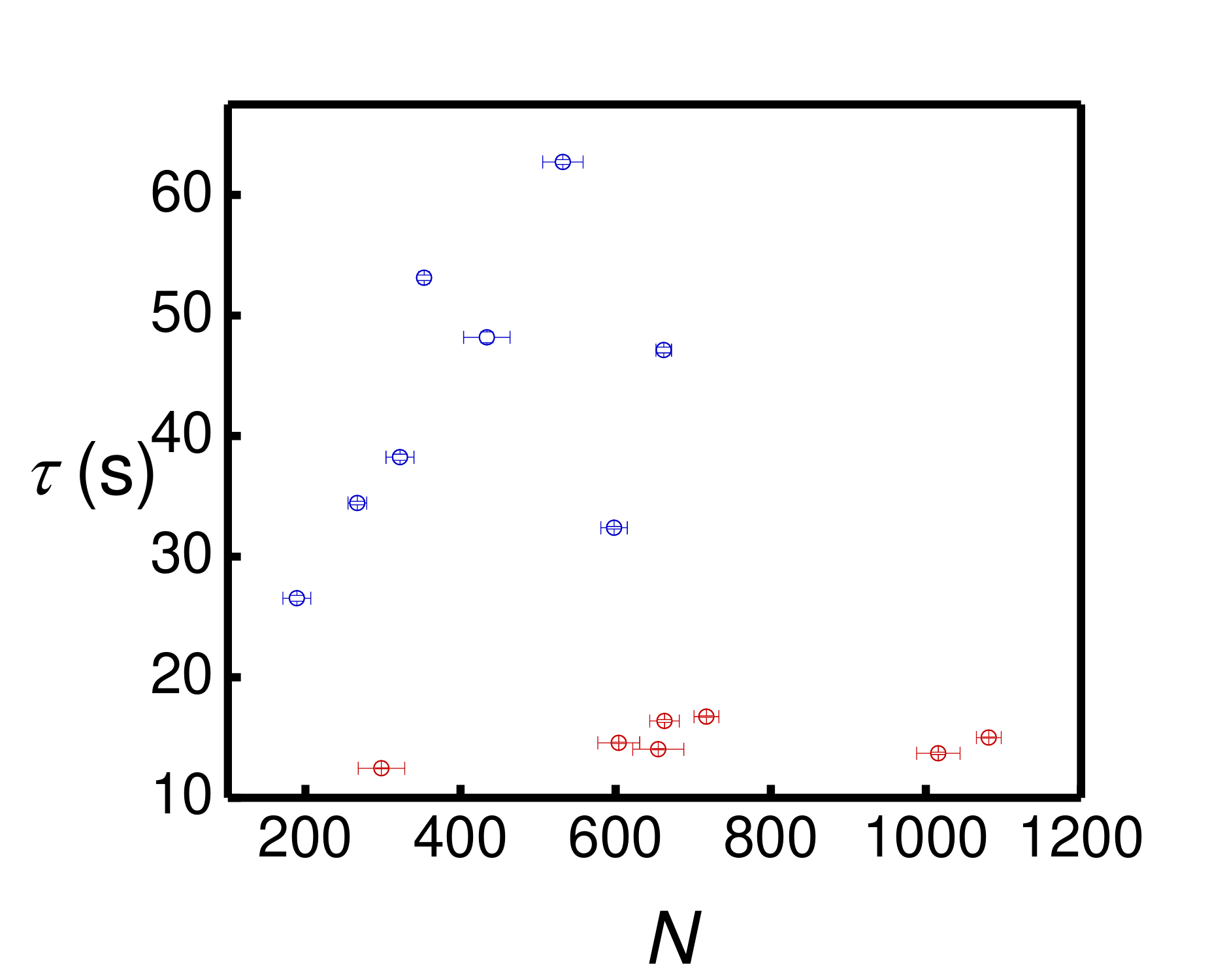}
\end{center}
\bfcaption{Particle turnover time depends on cluster size and particle speed}{ The turnover time $\tau$, defined in \cref{Fig retention}, depends on the number of particles $N$ in the cluster, and on particle speed and interactions. The number of particles is averaged over the time used to calculate the turnover time $\tau$ as in \cref{Fig retention}. This averaging time varies from $30$ to $180$ s in different clusters. Error bars are S.D. Blue points correspond to an applied electric field with amplitude $66$ V/mm and frequency $30$ kHz, producing an average single-particle speed of $12$ $\mu$m/s within clusters. Red points correspond to a field with the same frequency but amplitude $83$ V/mm, giving an average single-particle speed of $22$ $\mu$m/s within clusters. Particles with higher speeds and stronger interactions turn over more quickly.} \label{Fig turnover-size}
\end{figure*}

\begin{figure*}[!htb]
\begin{center}
\includegraphics[width=0.75\textwidth]{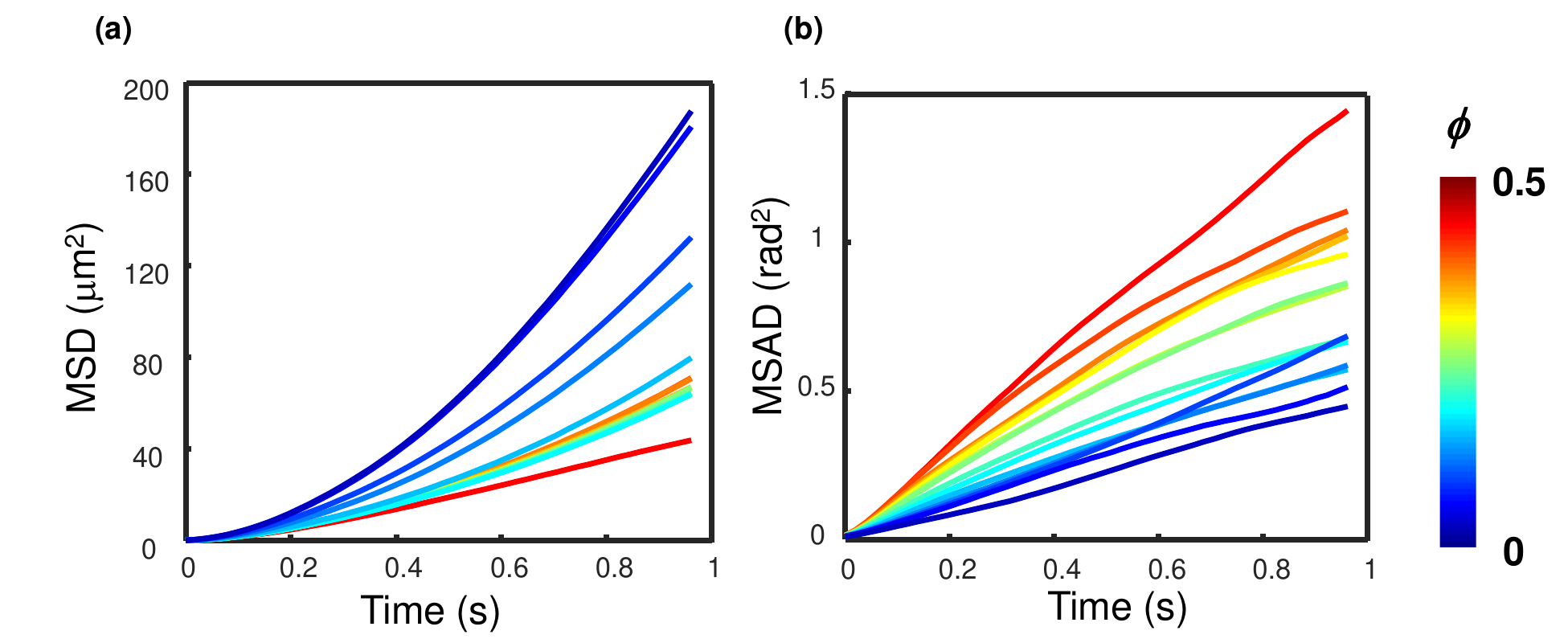}
\end{center}
  {\phantomsubcaption\label{Fig MSD-position}}
  {\phantomsubcaption\label{Fig MSD-angle}}
\bfcaption{Mean square displacements}{ Mean square displacements of particle position (\subref*{Fig MSD-position}) and angle (\subref*{Fig MSD-angle}) for particles at different local area fractions $\phi$. The particle speed and effective rotational diffusivity shown in \cref{Fig speed-rotation} are obtained from the first 0.5 s of these data by fitting $\textrm{MSD} = (v t)^2$ and $\textrm{MSAD} = D_{\text{r}}^{\text{eff}} t$, respectively.} \label{Fig MSD}
\end{figure*}

\begin{figure*}[!htb]
\begin{center}
\includegraphics[width=0.4\textwidth]{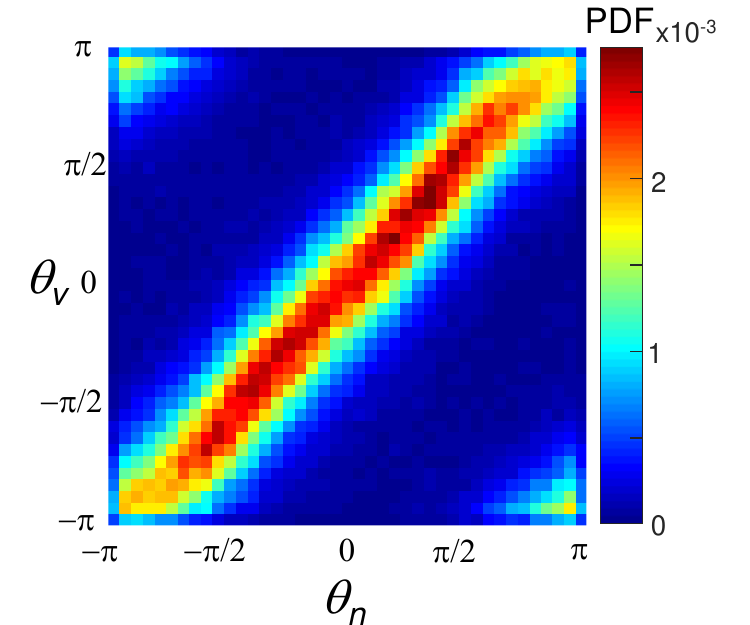}
\end{center}
\bfcaption{Orientation-velocity coincidence prior to clustering}{ Joint probability distribution function (PDF) of the particle orientation and velocity angles, as defined in \cref{Fig angles-definitions}, prior to cluster formation. In contrast to the mismatch found in clusters (\cref{Fig angles-histograms,Fig angles-joint-histogram}), the orientation and velocity directions coincide before clusters form.} \label{Fig angles-joint-histogram-uniform}
\end{figure*}

\begin{figure*}[!htb]
\begin{center}
\includegraphics[width=0.5\textwidth]{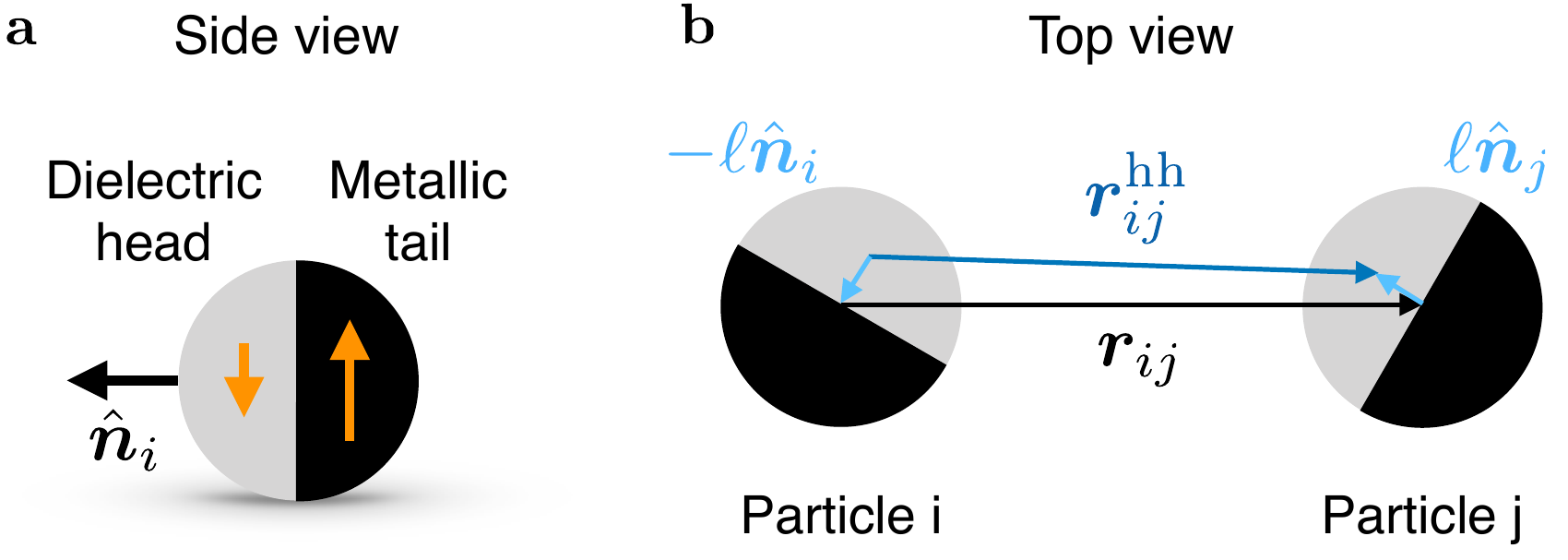}
\end{center}
  {\phantomsubcaption\label{Fig Janus-a}}
  {\phantomsubcaption\label{Fig Janus-b}}
\bfcaption{Schematic of Janus particles' polarization and dipolar interactions}{ \subref*{Fig Janus-a}, In the parameter regime used in the experiments, both hemispheres of the Janus particles polarize along the axis of the external electric field, parallel to the particle's equator. However, the metallic and dielectric hemispheres acquire effective dipolar moments of different magnitude and sign (orange arrows). The particles self-propel in a direction $\hat{\bm{n}}_i$ (black arrow), which points from the metallic tail to the dielectric head of the particles. \subref*{Fig Janus-b}, The net interaction between two particles results from all the interactions between their off-centered head and tail dipoles. For example, the interaction between the dipoles in the head hemispheres of two particles depends on their distance vector $\bm{r}_{ij}^{\text{hh}}$ (dark blue arrow). The schematic shows how this vector is related to the center-of-mass distance vector $\bm{r}_{ij}$ (black arrow): $\bm{r}_{ij}^{\text{hh}} = -\ell \hat{\bm{n}}_i + \bm{r}_{ij} + \ell \hat{\bm{n}}_j$. Light blue arrows represent the distance vector between the center of mass of the particles and the center of their head hemisphere.} \label{Fig Janus}
\end{figure*}

\begin{figure*}[!htb]
\begin{center}
\includegraphics[width=0.8\textwidth]{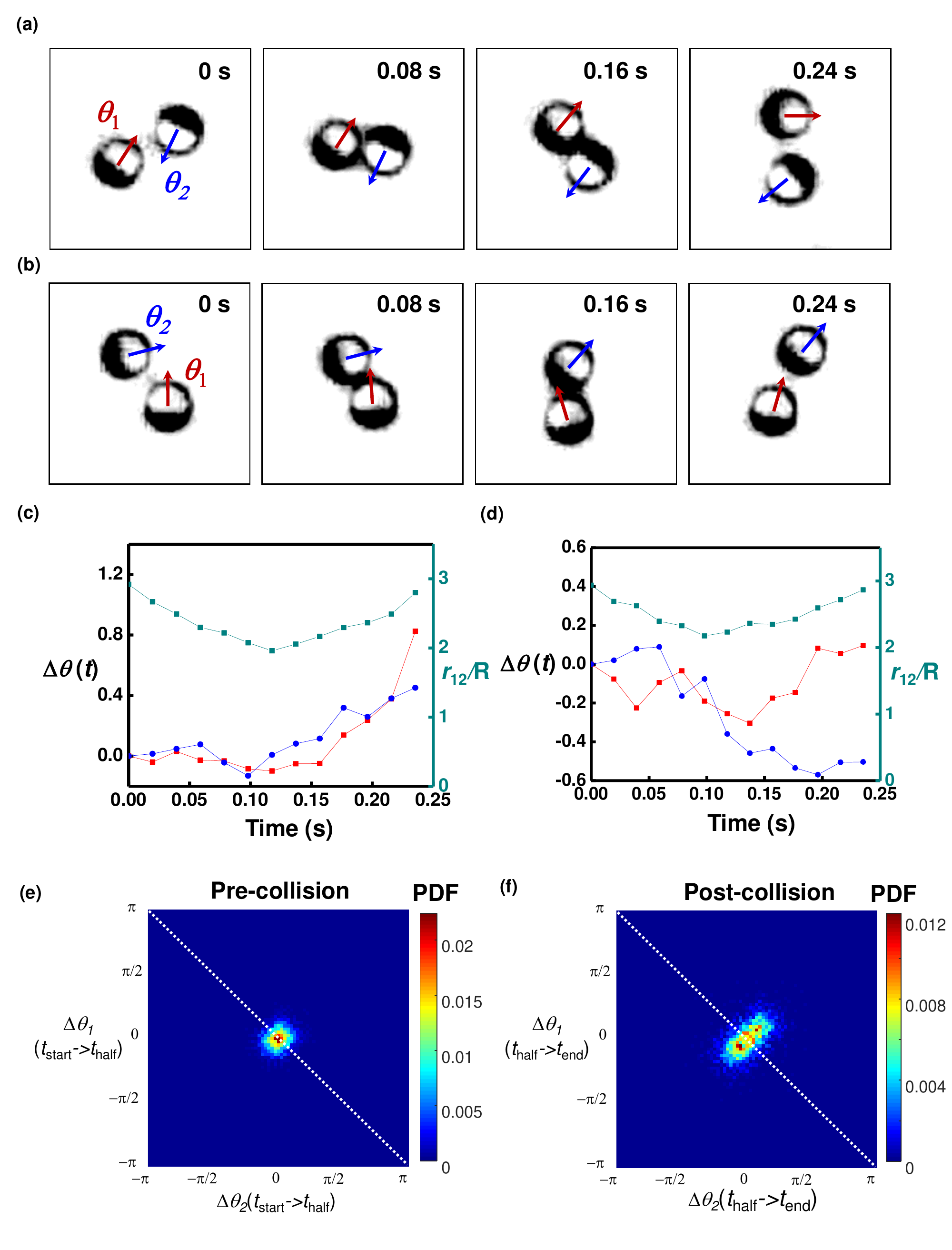}
\end{center}
  {\phantomsubcaption\label{Fig interaction-snapshot-1}}
  {\phantomsubcaption\label{Fig interaction-snapshot-2}}
  {\phantomsubcaption\label{Fig interaction-dynamics-1}}
  {\phantomsubcaption\label{Fig interaction-dynamics-2}}
  {\phantomsubcaption\label{Fig approach-statistics}}
  {\phantomsubcaption\label{Fig departure-statistics}}
\bfcaption{Dynamics and statistics of two-particle interaction events}{ \subref*{Fig interaction-snapshot-1},\subref*{Fig interaction-snapshot-2}, Two examples of interaction events between two particles, which we define by the condition $r_{12}< 3R$, with $R$ the particle radius. We also require the minimal interparticle distance to be $\min r_{12}< 2.2R$. We analyzed 2061 of such interaction events. In \subref*{Fig interaction-snapshot-1}, both particles turn clockwise. Therefore, the interaction torque has the same sign on both particles, showing that torques are non-reciprocal ($\bm{\Gamma}_{12} \neq -\bm{\Gamma}_{21}$). This type of interaction with particles initially pointing in opposite directions, defined by the condition $|\theta_1(t=0) + \theta_2(t=0)|<0.2$ rad, occurred in $\sim 33\%$ of the analyzed events. In \subref*{Fig interaction-snapshot-2}, particle 1 changes its orientation very little compared to particle 2, showing another example of non-reciprocal torques. The particles end up aligned in a chain. This type of interaction, defined by the condition $\min(\theta_1(t) - \theta_2(t))<0.5$ rad during the interaction event, occurred in $\sim 2.8\%$ of the analyzed events. \subref*{Fig interaction-dynamics-1},\subref*{Fig interaction-dynamics-2}, Evolution of the interparticle distance, $r_{12}$, and the change in angle of each particle, $\Delta \theta_{1,2}$, for the interactions events in \subref*{Fig interaction-snapshot-1} and \subref*{Fig interaction-snapshot-2}, respectively. \subref*{Fig approach-statistics},\subref*{Fig departure-statistics}, Joint probability distribution functions of the angle changes of each particle in an interacting pair. The pre-collision (\subref*{Fig approach-statistics}) and post-collision (\subref*{Fig departure-statistics}) phases respectively correspond to the times before and after the particles reach their minimal distance. These histograms show that, statistically, both particles in the interacting pair tend to turn in the same direction, showing that interaction torques are non-reciprocal. Reciprocal torques would lead to particles rotating in opposite directions and by the same magnitude, as indicated by the dashed lines.} \label{Fig interaction-statistics}
\end{figure*}

\begin{figure}[tb!]
\begin{center}
\includegraphics[width=0.75\textwidth]{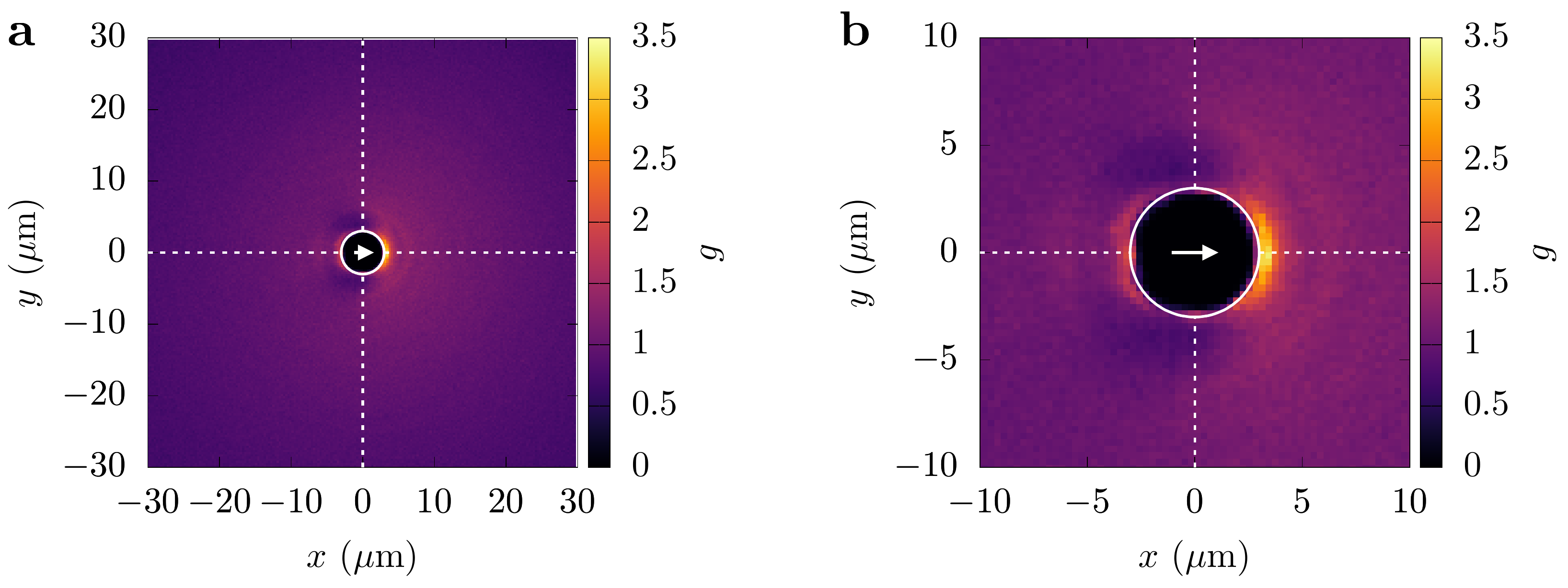}
\end{center}
  {\phantomsubcaption\label{fig gr-full}}
  {\phantomsubcaption\label{fig gr-zoom}}
\bfcaption{Pair distribution function measured in experiments}{ \subref*{fig gr-full}, Full $g(x,y)$ used in the calculation of the collective force and torque coefficients $\zeta_0$, $\zeta_1$, and $\tau_1$ (see text and \cref{eq coefficients}). \subref*{fig gr-zoom}, Zoomed-in region, which allows to more clearly appreciate that the pair distribution function is anisotropic, indicating that it is more likely to find another particle in front than behind a reference self-propelled particle. In each panel, the arrow indicates the direction of self-propulsion of the reference particle, dashed lines indicate the coordinate axes, and the white circle indicates the region $r<2R$ of volume exclusion between two particles.} \label{Fig gr}
\end{figure}

\begin{figure*}[!htb]
\begin{center}
\includegraphics[width=0.75\textwidth]{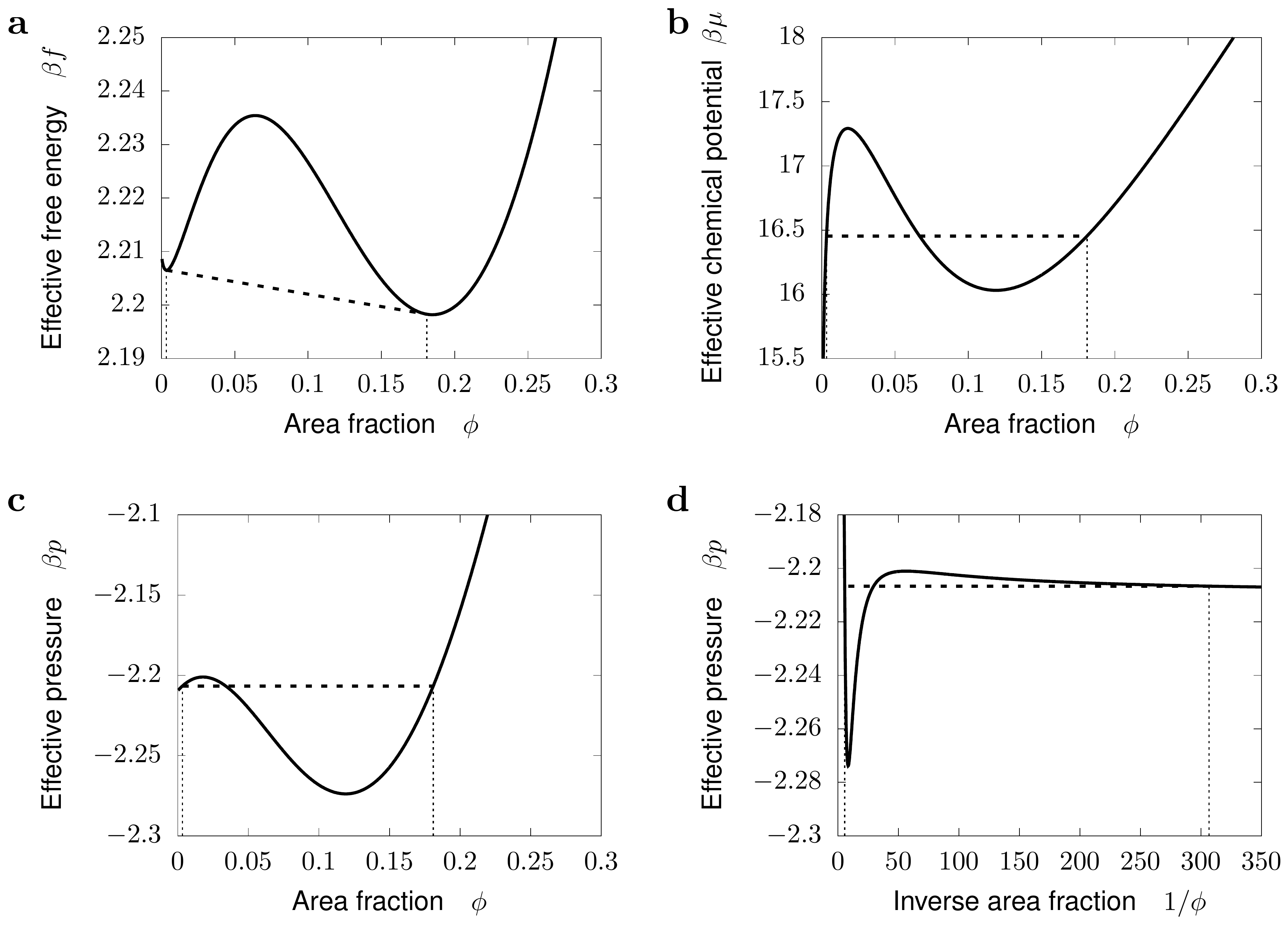}
\end{center}
  {\phantomsubcaption\label{fig free-energy}}
  {\phantomsubcaption\label{fig chemical-potential}}
  {\phantomsubcaption\label{fig pressure}}
  {\phantomsubcaption\label{fig pressure-Maxwell}}
\bfcaption{Effective thermodynamics of torque-based MIPS}{ Effective free energy density (\subref*{fig free-energy}, \cref{eq free-energy}), chemical potential (\subref*{fig chemical-potential}, \cref{eq chemical-potential}), and thermodynamic pressure (\subref*{fig pressure}-\subref*{fig pressure-Maxwell}, \cref{eq pressure}) of the active Janus suspension as a function of the area fraction of particles, for $v_0=10$ $\mu$m/s. The remaining parameter values are evaluated using the estimates in \cref{t parameters}. To better visualize the double-well shape of the free energy $\beta f(\phi)$, we added a linear term $-16.5\phi$, which does not affect phase coexistence. Thin dashed lines indicate the densities of the coexisting phases. These densities are obtained from the common-tangent construction on the free energy density (\subref*{fig free-energy}), which corresponds to equality of both chemical potential (\subref*{fig chemical-potential}) and pressure (\subref*{fig pressure}), as indicated by thick dashed lines. Alternatively, the common-tangent construction also corresponds to the Maxwell construction on the curve $p(1/\phi)$, as indicated by the thick dashed line in panel \subref*{fig pressure-Maxwell}.} \label{Fig thermodynamics}
\end{figure*}

\begin{figure*}[!htb]
\begin{center}
\includegraphics[width=0.7\textwidth]{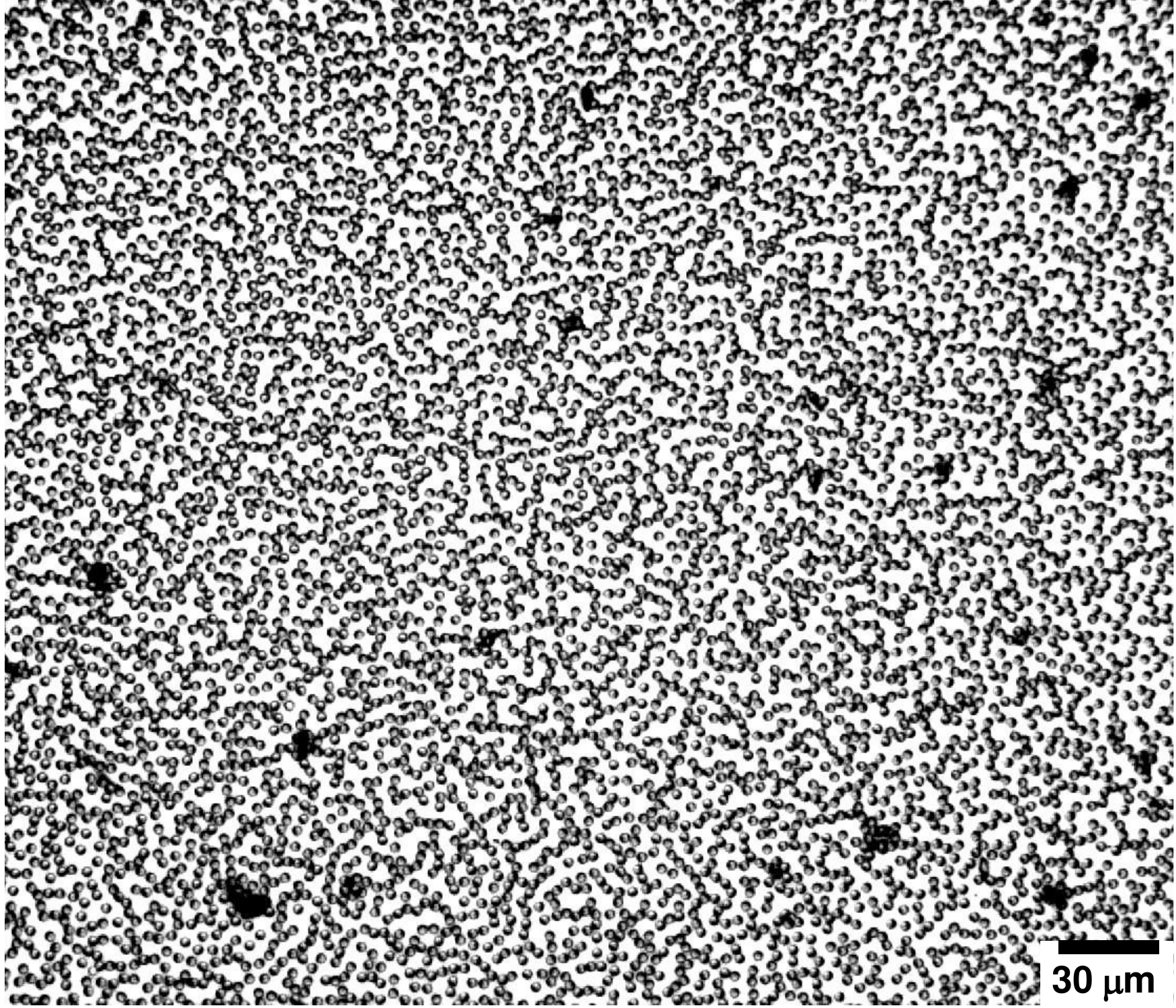}
\end{center}
\bfcaption{High-density uniform phase}{ Snapshot of the uniform phase found at $\phi = 0.35$, showing transient particle chains throughout the system.} \label{Fig high-density-uniform}
\end{figure*}

\clearpage

\section*{Supplementary Movies} \label{movies}

\noindent\textbf{Supplementary Movie 1: Coarsening dynamics during active phase separation.} This movie, taken with a 5x objective, shows the phase separation process of a quasi-2D Janus particle system (area fraction $\phi \sim 0.12$) when AC electric field ($30$ kHz, $83$ V/mm) perpendicular to the imaging plane is applied. The movie is played at 10x real speed. 

\bigskip

\noindent\textbf{Supplementary Movie 2: Growth of a representative active cluster.} This movie, from which the snapshots in \cref{Fig cluster-growth} are taken, is obtained using a perpendicular AC electric field ($30$ kHz, $83$ V/mm), and is played at 20x real speed. 

\bigskip

\noindent\textbf{Supplementary Movie 3: Particle turnover dynamics for an active cluster whose population remains nearly constant.} The movie is taken after the perpendicular AC electric field ($30$ kHz, $67$ V/mm) has been applied for a few minutes and the clusters are coarsening. Particles are color-coded according to their distance from the cluster centroid in the first frame, which is an arbitrary time point. Particles joining the cluster after the first frame are not labeled. This movie is played at 2x real speed and associated with \cref{Fig 2}.

\bigskip

\noindent\textbf{Supplementary Movie 4: Particle turnover dynamics for two clusters of different sizes.} The plots on the right refer to the movies on the left. Native particles are defined as those particles within the frame at the beginning of the movie. $N_0(t)$ is the number of native particles remaining in the frame after time $t$. Both movies are taken after the perpendicular AC electric field ($30$ kHz, $67$ V/mm) has been applied for a few minutes. Particles are color-coded according to their distance from the cluster centroid in the first frame, which is an arbitrary time point. Particles joining the cluster after the first frame are not labeled. This movie is played at 2x real speed.

\bigskip

\noindent\textbf{Supplementary Movie 5: Flickering chains in an active cluster.} Lines are drawn to connect pairs of particles whose centers are separated by less than $3R$ ($R=1.5$ $\mu$m is the particle radius), color-coded by the degree of pair alignment, with red indicating alignment and blue anti-alignment. This movie is played at 0.2x real speed. A snapshot of this movie is presented in \cref{Fig orientations}.

\bigskip

\noindent\textbf{Supplementary Movie 6: High-density uniform phase with transient particle chains.} This movie, taken with a 20x objective, shows the high-density uniform phase (area fraction $\phi\sim 0.35$) with transient particle chains when AC electric field ($30$ kHz, $83$ V/mm) perpendicular to the imaging plane is applied. A snapshot of this movie is presented in \cref{Fig high-density-uniform}.

\clearpage

\twocolumngrid

\section*{Supplementary Note} \label{theory}

In this Supplementary Note, we provide a detailed account of our theory for torque-based motility-induced phase separation. We start by proposing a simple microscopic model for the dynamics of self-propelled Janus colloids interacting via electrostatic forces. Because these forces are non-central, they lead to interparticle torques. We then systematically coarse-grain the microscopic equations of motion to obtain a hydrodynamic description in terms of continuum equations, which describe the collective behavior of the active colloids. This coarse-graining derivation shows how interparticle torques reorient particle motion toward regions with higher particle density. We then show that, when combined with particle self-propulsion, this reorientation produces a spinodal instability leading to phase separation. We predict the phase diagram, including the densities of the coexisting dense and dilute phases. Finally, we provide experimental estimates for the model parameters. Overall, we demonstrate a new mechanism for motility-induced phase separation (MIPS) that relies on interparticle torques instead of central forces. Thus, this torque-based MIPS mechanism is an alternative to the originally-proposed one, namely the slowdown of particle motion with increasing density due to repulsion between the particles \cite{Cates2015,Speck2020}.

\subsection{Microscopic model: Active Janus particles with electrostatic interactions} \label[SI section]{microscopic-model}

\subsubsection{Dielectric response and electrostatic interactions between Janus colloids} \label[SI section]{electrodynamics}

In our experiments, metal-dielectric Janus particles are driven by an alternating electric field. This field induces an electric polarization in each hemisphere of the Janus particles (\cref{Fig Janus-a}). Due both to Ohmic losses and to the Maxwell-Wagner effect produced by the metal coating on one of the hemispheres, the electric polarization has a delay with respect to the electric field \cite{Jones1995}. This dynamic response can be taken into account via a complex and frequency-dependent dielectric permittivity of the material
\begin{equation} \label{eq complex-permittivity}
\bar{\epsilon}(\omega)= \epsilon - i \frac{\sigma}{\omega},
\end{equation}
where $\epsilon$ is the static dielectric permittivity, $\sigma$ is the conductivity of the material, and $\omega$ is the angular frequency.

The electric polarization of the hemispheres results in electrostatic interactions between the particles. Here, we model these interactions as resulting from point dipoles located at the center of each hemisphere. Taking into account electrodynamic effects via the complex permittivity in \cref{eq complex-permittivity}, the dipole moment induced by an alternating electric field with amplitude $\bm{E}_0$ and frequency $\nu$ in a hemisphere of radius $R$ with isotropic properties is given by \cite{Jones1995}
\begin{equation}
\bm{d}_{\text{c}}(\nu) = 2\pi\epsilon K(\nu) R^3 \bm{E}_0,
\end{equation}
where the subscript c stands for complex. Here, $K(\nu)$ is the Clausius-Mossotti function that characterizes the frequency dependence, which was previously measured for each hemisphere of our Janus particles \cite{Yan2016}.

In our particles, electric dipoles are either parallel or antiparallel to each other, and they are perpendicular to the particles' plane of motion (\cref{Fig Janus-a}). For this specific arrangement, the electrostatic force exerted by a complex point dipole a on another complex point dipole b is given by \cite{Jones1995}
\begin{equation}
\bm{F}_{\text{ab}} = \frac{3\,\mathrm{Re}[\bm{d}_{\text{c,a}}^* \cdot\bm{d}_{\text{c,b}}]}{4\pi\epsilon r^4} \hat{\bm{r}},
\end{equation}
where $\bm{r} = \bm{r}_{\text{b}} - \bm{r}_{\text{a}}$ is the distance vector between the two dipoles, and $r=|\bm{r}|$. This interaction can be recast as
\begin{equation} \label{eq dipolar-force}
\bm{F}_{\text{ab}} = \frac{3 \bm{d}_{\text{a}} \cdot \bm{d}_{\text{b}}}{4\pi \epsilon r^4} \hat{\bm{r}},
\end{equation}
where $\bm{d}_{\text{a}}$ and $\bm{d}_{\text{b}}$ are effective non-complex dipoles. The effective dipoles $d_{\text{h}}$ and $d_{\text{t}}$ of the head and tail hemispheres of our Janus particles, respectively, are
\begin{equation}
d_{\text{h},\text{t}} = 2\pi \epsilon R^3 E_0 |K_{\text{h},\text{t}}(\nu)|,
\end{equation}
where $|K_{\text{h},\text{t}}(\nu)| = \sqrt{\mathrm{Re}[K_{\text{h},\text{t}}(\nu)]^2 + \mathrm{Im}[K_{\text{h},\text{t}}(\nu)]^2}$ are the moduli of the complex numbers $K_{\text{h},\text{t}}(\nu)$. Moreover, whereas the squares of the effective dipole moments are directly the $d_{\text{h}}^2$ and $d_{\text{t}}^2$, the product $d_t d_h$ is given by
\begin{equation}
d_{\text{h}} d_{\text{t}} = (2\pi\epsilon R^3 E_0)^2 \mathrm{Re}[K_{\text{h}}^* K_{\text{t}}].
\end{equation}
With these substitutions, we can encode the dynamic dielectric response of the particles into effective dipole moments.

\subsubsection{Electrostatic force and torque between Janus colloids} \label[SI section]{force-torque}

The dipole-dipole forces described by \cref{eq dipolar-force} are isotropic and central. However, the dipoles are located at the center of each hemisphere, and not at the center of the particle (\cref{Fig Janus-a}). Moreover, the effective dipole in the metallic hemisphere (particle tail, $d_{\text{t}}$) is different in magnitude and sign than the dipole in the dielectric hemisphere (particle head, $d_{\text{h}}$). As a result, the dipolar forces on the head and tail hemispheres are different. Thus, the electrostatic interactions between two particles are non-central, and they produce not only a net force but also a net torque (\cref{Fig interactions}). We obtain both the net force and the net torque below.

For particle $i$, the dipoles of the head and tail hemispheres are respectively located at $\bm{r}_i^{\text{h,t}} = \pm\ell\hat{\bm{n}}_i$, where $\hat{\bm{n}}_i$ is the direction of self-propulsion, perpendicular to the equator (\cref{Fig Janus}), and $\ell = 3R/8$ is the distance between the center of mass of the hemisphere and that of the particle. The total force exerted by particle $i$ on particle $j$ is the sum of four dipolar contributions (head-head, head-tail, tail-head, and tail-tail):
\begin{equation}
\bm{F}_{ij} = \bm{F}_{ij}^{\text{hh}} + \bm{F}_{ij}^{\text{ht}} + \bm{F}_{ij}^{\text{th}} + \bm{F}_{ij}^{\text{tt}}.
\end{equation}
Each term involves its own dipolar moments and distance vector. The head and tail dipolar moments are $\bm{d}_{\text{h}} = d_{\text{h}}\hat{\bm{z}}$ and $\bm{d}_{\text{t}} = d_{\text{t}}\hat{\bm{z}}$, with $d_{\text{t}}>0$, $d_{\text{h}}<0$, and $d_{\text{t}}>|d_{\text{h}}|$ (\cref{Fig Janus-a}). Respectively, the distance vectors for each contribution can be written in terms of the distance vector $\bm{r}_{ij}$ between the particles' centers of mass and their individual self-propulsion directions $\hat{\bm{n}}_i$ and $\hat{\bm{n}}_j$. Performing the vector sums illustrated in \cref{Fig Janus-b}, the distance vectors for each dipolar interaction are
\begin{subequations}
\begin{align}
\bm{r}_{ij}^{\text{hh}} &= -\ell \hat{\bm{n}}_i + \bm{r}_{ij} + \ell \hat{\bm{n}}_j,\\
\bm{r}_{ij}^{\text{ht}} &= -\ell \hat{\bm{n}}_i + \bm{r}_{ij} - \ell \hat{\bm{n}}_j,\\
\bm{r}_{ij}^{\text{th}} &= \ell \hat{\bm{n}}_i + \bm{r}_{ij} + \ell \hat{\bm{n}}_j,\\
\bm{r}_{ij}^{\text{tt}} &= \ell \hat{\bm{n}}_i + \bm{r}_{ij} - \ell \hat{\bm{n}}_j.
\end{align}
\end{subequations}
In terms of these vectors, the net interparticle force is expressed as
\begin{equation}
\bm{F}_{ij} = \frac{3}{4\pi\epsilon} \left[\frac{d_h^2}{\left|\bm{r}_{ij}^{\text{hh}}\right|^4} \hat{\bm{r}}_{ij}^{\text{hh}} + \frac{d_h d_t}{\left|\bm{r}_{ij}^{\text{ht}}\right|^4} \hat{\bm{r}}_{ij}^{\text{ht}} + \frac{d_t d_h}{\left|\bm{r}_{ij}^{\text{th}}\right|^4} \hat{\bm{r}}_{ij}^{\text{th}} + \frac{d_t^2}{\left|\bm{r}_{ij}^{\text{tt}}\right|^4} \hat{\bm{r}}_{ij}^{\text{tt}}\right].
\end{equation}
Note that this force is non-central; in addition to a component along the interparticle distance axis $\hat{\bm{r}}_{ij}$, it has components along each particle's self-propulsion direction $\hat{\bm{n}}_{i}$ and $\hat{\bm{n}}_{j}$.

The off-centered dipolar interactions lead to torques. Like the net force, the total torque exerted by particle $i$ on the center of mass of particle $j$ also has four contributions:
\begin{equation} \label{eq total-force}
\bm{\Gamma}_{ij} = \bm{r}_j^{\text{h}} \times \left(\bm{F}_{ij}^{\text{hh}} + \bm{F}_{ij}^{\text{th}}\right) + \bm{r}_j^{\text{t}} \times \left(\bm{F}_{ij}^{\text{ht}} + \bm{F}_{ij}^{\text{tt}}\right).
\end{equation}
Using the distance vectors between the head and tail hemispheres, the center-of-mass torque can be expressed as
\begin{multline} \label{eq total-torque}
\bm{\Gamma}_{ij} = \frac{3\ell}{4\pi\epsilon}\left[\left(\frac{d_h^2}{\left|\bm{r}_{ij}^{\text{hh}}\right|^5} + \frac{d_t d_h}{\left|\bm{r}_{ij}^{\text{th}}\right|^5} -  \frac{d_h d_t}{\left|\bm{r}_{ij}^{\text{ht}}\right|^5} -  \frac{d_t^2}{\left|\bm{r}_{ij}^{\text{tt}}\right|^5}\right)\hat{\bm{n}}_j\times \bm{r}_{ij} \right.\\
\left. + \left(- \frac{d_h^2}{\left|\bm{r}_{ij}^{\text{hh}}\right|^5} +  \frac{d_t d_h}{\left|\bm{r}_{ij}^{\text{th}}\right|^5} +  \frac{d_h d_t}{\left|\bm{r}_{ij}^{\text{ht}}\right|^5} -  \frac{d_t^2}{\left|\bm{r}_{ij}^{\text{tt}}\right|^5}\right) \ell \hat{\bm{n}}_j\times \hat{\bm{n}}_i\right].
\end{multline}
Similar to the net force, the center-of-mass torque has a term involving the interparticle distance vector $\hat{\bm{r}}_{ij}$ but also a term associated entirely with the particles' self-propulsion direction, $\hat{\bm{n}}_i$ and $\hat{\bm{n}}_j$.

Throughout most of the experiments, and certainly in the initial condition, the interparticle distance $r_{ij}$ is typically much larger than the distance $\ell \approx 0.5$ $\mu$m between the center of mass of the particle and of either hemisphere, $r_{ij}\gg \ell$. We leverage this condition to obtain simpler expressions for the interparticle force and torque. Expanding \cref{eq total-force,eq total-torque} to lowest order in $\ell/r_{ij}$, we obtain
\begin{equation}
\bm{F}_{ij} \approx \frac{3}{4\pi\epsilon} \frac{(d_{\text{h}} + d_{\text{t}})^2}{r_{ij}^4} \hat{\bm{r}}_{ij},
\end{equation}
\begin{equation}
\bm{\Gamma}_{ij} \approx \frac{3\ell}{4\pi\epsilon} \frac{d_{\text{h}}^2 - d_{\text{t}}^2}{r_{ij}^4}\hat{\bm{n}}_j\times \hat{\bm{r}}_{ij}.
\end{equation}
At this level of approximation, the net interparticle force $\bm{F}_{ij}$ is central (i.e., along $\hat{\bm{r}}_{ij}$). Respectively, given that $d_{\text{t}}^2>d_{\text{h}}^2$, the net interparticle torque $\bm{\Gamma}_{ij}$ tends to rotate particles toward the direction of the interparticle distance (\cref{Fig interactions}).

Finally, the electrodes that generate the electric field are equipotential surfaces. Therefore, they screen the electrostatic interactions between particles over a length scale comparable to the distance between electrodes, $\lambda = 120$ $\mu$m (\cref{Fig system}). Without solving the full electrostatic problem in detail, we account for screening effects by adding an exponential factor $e^{-r_{ij}/\lambda}$. Thus, the final expressions of the interparticle force and torque are
\begin{subequations} \label{eq particle-interactions}
\begin{align}
\bm{F}_{ij} &\approx \frac{3}{4\pi\epsilon} \frac{(d_{\text{h}} + d_{\text{t}})^2}{r_{ij}^4} e^{-r_{ij}/\lambda}\; \hat{\bm{r}}_{ij}, \label{eq interaction-force}\\
\bm{\Gamma}_{ij} &\approx \frac{3\ell}{4\pi\epsilon} \frac{d_{\text{h}}^2 - d_{\text{t}}^2}{r_{ij}^4} e^{-r_{ij}/\lambda}\; \hat{\bm{n}}_j\times \hat{\bm{r}}_{ij}. \label{eq interaction-torque}
\end{align}
\end{subequations}

\subsubsection{Langevin equations of motion} \label[SI section]{Langevin}

Upon the application of an electric field, our metal-dielectric Janus particles become self-propelled by virtue of induced-charge electrophoresis \cite{Gangwal2008,Moran2017}. Therefore, in addition to the interaction force and torque in \cref{eq particle-interactions}, our Janus particles are driven by self-propulsion forces, as well as both translational and rotational fluctuations. These driving forces are balanced by damping forces in the form of viscous friction between the particles and the solvent. Hydrodynamic interactions between particles are negligible in front of electrostatic interactions \cite{Yan2016}. Putting all together, we can write the Langevin equations for the translational and rotational motion of particle $i$ as
\begin{subequations} \label{eq Langevin}
\begin{align}
\frac{\dd\bm{r}_i}{\dd t} &= v_0 \hat{\bm{n}}_i + \frac{\bm{F}_i}{\xi_{\text{t}}} + \bm{\eta}^{\text{t}}_i(t);\qquad \bm{F}_i = \sum_{j\neq i} \bm{F}_{ji}, \label{eq Langevin-translation}\\
\frac{\dd\hat{\bm{n}}_i}{\dd t} &= \frac{\bm{\Gamma}_i}{\xi_{\text{r}}} + \bm{\eta}_i^{\text{r}}(t);\qquad \bm{\Gamma}_i= \sum_{j\neq i} \bm{\Gamma}_{ji}. \label{eq Langevin-rotation}
\end{align}
\end{subequations}
Here, $v_0$ is the self-propulsion speed, and $\xi_{\text{t}}$ and $\xi_{\text{r}}$ are the translational and rotational friction coefficients, respectively, which we assume to be isotropic. Finally, $\bm{\eta}_i^{\text{t}}(t)$ and $\bm{\eta}_i^{\text{r}}(t)$ are respectively translational and rotational Gaussian white noises with zero mean and correlations given by
\begin{subequations} \label{eq noises}
\begin{align}
\left\langle \eta^{\text{t}}_{i,\alpha}(t)\eta^{\text{t}}_{j,\beta}(t')\right\rangle &= 2D_{\text{t}}\, \delta_{ij} \delta_{\alpha\beta} \delta(t-t'), \label{eq translational-noise}\\
\left\langle \eta^{\text{r}}_{i,\alpha}(t)\eta^{\text{r}}_{j,\beta}(t')\right\rangle &= 2D_{\text{r}}\, \delta_{ij}\delta_{\alpha\beta}\delta(t-t'). \label{eq rotational-noise}
\end{align}
\end{subequations}
Here, Greek indices indicate spatial components. Respectively, $D_{\text{t}}$ and $D_{\text{r}}$ are the bare translational and rotational diffusion coefficients of the particles (in the absence of self-propulsion), which, as for the friction coefficients, we assume to be isotropic.

\subsubsection{Torque on the common center of mass of two particles} \label[SI section]{relative-torque}

Before proceeding to coarse-grain this microscopic model, we note that the interaction torques $\bm{\Gamma}_{ij}$ are non-reciprocal, i.e. $\bm{\Gamma}_{ij} \neq -\bm{\Gamma}_{ji}$. These torques act on the center of mass of each particle in the interacting pair, affecting particle orientation. To ensure overall torque balance, these interaction torques combine with the relative torque $\bm{\Gamma}_{\text{rel}}$ of the particles with respect to their common center of mass, located halfway between the two particles.
To obtain $\bm{\Gamma}_{\text{rel}}$, we obtain the corresponding angular momentum:
\begin{equation} \label{eq angular-momentum}
\bm{L}_{\text{rel}} = m \frac{\bm{r}_{ij}}{2}\times \bm{v}_j + m \frac{\bm{r}_{ji}}{2} \times \bm{v}_j = \frac{m}{2} \bm{r}_{ij} \times (\bm{v}_j - \bm{v}_i),
\end{equation}
where $m$ is the particle mass. Ignoring noise, we introduce the particle velocities using \cref{eq Langevin-translation} to obtain
\begin{equation} \label{eq relative}
\bm{L}_{\text{rel}} = \frac{m}{2} v_0\; \bm{r}_{ij} \times (\hat{\bm{n}}_j - \hat{\bm{n}}_i).
\end{equation}
We then obtain the relative torque as
\begin{multline}
\bm{\Gamma}_{\text{rel}} = \frac{\dd \bm{L}_{\text{rel}}}{\dd t} = \frac{m}{2} v_0 \left[ (\bm{v}_j - \bm{v}_i) \times (\hat{\bm{n}}_j - \hat{\bm{n}}_i) \phantom{\frac{\dd \hat{\bm{n}}_j}{\dd t}}\right.\\
\left.+ \bm{r}_{ij}\times \left( \frac{\dd \hat{\bm{n}}_j}{\dd t} - \frac{\dd \hat{\bm{n}}_i}{\dd t}\right)\right].
\end{multline}
Using \cref{eq Langevin}, ignoring noise, we obtain
\begin{equation}
\bm{\Gamma}_{\text{rel}} = m v_0 \left[ \frac{1}{\xi_{\text{t}}} \bm{F}_{ij} \times (\hat{\bm{n}}_j - \hat{\bm{n}}_i) + \frac{1}{2 \xi_{\text{r}}} \bm{r}_{ij}\times (\bm{\Gamma}_{ij} - \bm{\Gamma}_{ji}) \right],
\end{equation}
where we have used that interaction forces are reciprocal but torques are not.

\subsection{Coarse-graining: From the microscopic model to a hydrodynamic description} \label[SI section]{coarse-graining}

In this section, we systematically coarse-grain the microscopic equations of motion \cref{eq Langevin} to derive hydrodynamic equations that capture the collective behavior of the active Janus particle suspension. To this end, we use standard methods of non-equilibrium statistical mechanics \cite{Zwanzig2001,Mazo2002,Balescu1975,LeBellac2004}, performing the derivation in steps:
\begin{enumerate}
\item We go from the Langevin equations of motion for the set of $N$ interacting particles to the Smoluchowski equation, which describes the evolution of the system in terms of the $N$-particle distribution function. We then project the Smoluchowski equation onto particle coordinates to obtain the Bogoliubov-Born-Green-Kirkwood-Yvon (BBGKY) hierarchy of equations for the 1,2,3,...-particle distribution functions. We truncate this hierarchy at the 2-particle order. In this way, information about pair correlations becomes encoded into two interaction integrals known as the collective force and torque, respectively \cite{Kirkwood1949,Irving1950,Speck2020}.
\item We perform a gradient expansion to obtain the collective force and torque in terms of the particle density field and its gradient.
\item We define continuum fields as moments of the one-particle distribution function, and obtain a hierarchy of hydrodynamic equations for these fields. Finally, we truncate the hierarchy at second order to obtain closed hydrodynamic equations to describe the large-scale behavior of our system.
\end{enumerate}
We provide the details for each of these steps in separate subsections below.

\subsubsection{Smoluchowski equation and BBGKY hierarchy} \label[SI section]{Smoluchowski}

The behavior of the system encoded in the set of coupled Langevin equations \cref{eq Langevin} can be equivalently described by the Smoluchowski equation, i.e. the Fokker-Planck equation for the N-particle distribution function $\Psi_N(\bm{r}_1,\hat{\bm{n}}_1,\ldots,\bm{r}_N,\hat{\bm{n}}_N;t)$:
\begin{equation} \label{eq N-particle}
\partial_t \Psi_N = - \sum_{i=1}^N \left[ \bm{\nabla}_i \cdot \bm{J}_{\text{t},i} + \hat{\bm{n}}_i\times \partial_{\hat{\bm{n}}_i} \cdot \bm{J}_{\text{r},i}\right].
\end{equation}
Here, $\hat{\bm{n}}\times \partial_{\hat{\bm{n}}}$ is the rotation operator \cite{Marchetti2013}, and $\bm{J}_{\text{t}}$ and $\bm{J}_{\text{r}}$ are the translational and rotational probability currents, respectively, given by
\begin{subequations}
\begin{align}
\bm{J}_{\text{t},i} &= \left[ v_0\hat{\bm{n}}_i + \frac{\bm{F}_i}{\xi_{\text{t}}}\right] \Psi_N - D_{\text{t}}\,\bm{\nabla}_i\Psi_N,\\
\bm{J}_{\text{r},i} &= \frac{\bm{\Gamma}_i}{\xi_{\text{r}}}\Psi_N - D_{\text{r}}\,\hat{\bm{n}}_i\times \partial_{\hat{\bm{n}}_i}\Psi_N.
\end{align}
\end{subequations}
Hereafter, we describe the two-dimensional particle orientation $\hat{\bm{n}}_i$ in terms of the angle $\theta_i$: $\hat{\bm{n}}_i = (\cos\theta_i,\sin\theta_i)^T$.

Next, we obtain the BBGKY hierarchy of equations for the distribution functions of increasing order. The distribution function of order $k$ is defined as
\begin{multline} \label{eq k-particle}
\Psi_k(\bm{r}_1,\theta_1,\ldots,\bm{r}_k,\theta_k;t) \equiv \frac{N!}{(N-k)!} \int \dd^2\bm{r}_{k+1}\cdots \dd^2\bm{r}_N \\
\times \int \dd\theta_{k+1}\cdots\dd\theta_N\; \Psi_N(\bm{r}_1,\theta_1,\ldots,\bm{r}_N,\theta_N;t).
\end{multline}
We start with the one-particle distribution function $\Psi_1(\bm{r}_1,\theta_1;t)$, which is the probability density of finding one particle (tagged with label 1) at position $\bm{r}_1$, with orientation $\theta_1$, at time $t$, regardless of the positions and orientations of the other $N-1$ particles. Projecting \cref{eq N-particle} according to the definition in \cref{eq k-particle} for $k=1$, we obtain
\begin{multline} \label{eq psi1}
\partial_t \Psi_1 = -\bm{\nabla}_1\cdot\left[\left( v_0\hat{\bm{n}}_1 - D_{\text{t}}\bm{\nabla}_1\right) \Psi_1\right] - \bm{\nabla}_1 \cdot \frac{\bm{F}_{\text{int}}}{\xi_{\text{t}}} \\
 - \partial_{\theta_1} \frac{\hat{\bm{z}}\cdot\bm{\Gamma}_{\text{int}}}{\xi_{\text{r}}} + D_{\text{r}}\, \partial_{\theta_1}^2 \Psi_1.
\end{multline}
Here, $\bm{F}_{\text{int}}$ and $\bm{\Gamma}_{\text{int}}$ are the collective force and torque, respectively, which encode the effects of interactions on the tagged particle. For pair-wise interactions, these collective quantities can be expressed in terms of the two-particle distribution function $\Psi_2 (\bm{r}_1,\theta_1,\bm{r}_2,\theta_2;t)$ as
\begin{subequations} \label{eq collective-psi2}
\begin{align}
&\begin{multlined}
\bm{F}_{\text{int}}(\bm{r}_1,\theta_1;t) =\\
 -\int \dd^2\bm{r}' \; F(\left| \bm{r}'-\bm{r}_1\right|)\, \frac{\bm{r}'-\bm{r}_1}{\left| \bm{r}'-\bm{r}_1\right|} \,\Psi_2(\bm{r}_1,\theta_1,\bm{r}';t),
 \end{multlined}
\\
&\begin{multlined}
\bm{\Gamma}_{\text{int}}(\bm{r}_1,\theta_1;t) =\\
 \int \dd^2\bm{r}' \; \Gamma(\left| \bm{r}'-\bm{r}_1\right|)\, \hat{\bm{n}}_1 \times \frac{\bm{r}'-\bm{r}_1}{\left| \bm{r}'-\bm{r}_1\right|}\, \Psi_2(\bm{r}_1,\theta_1,\bm{r}';t),
\end{multlined}
\end{align}
\end{subequations}
where we have already integrated over the orientation $\theta_2$ of particle 2. Here,
\begin{subequations} \label{eq force-torque-fields-expressions}
\begin{align}
F(r) &= \frac{3(d_{\text{h}} + d_{\text{t}})^2}{4\pi\epsilon} \frac{e^{-r/\lambda}}{r^4}, \label{eq force-field}\\
\Gamma(r) &= \frac{3\ell(d_{\text{h}}^2 - d_{\text{t}}^2)}{4\pi\epsilon} \frac{e^{-r/\lambda}}{r^4} \label{eq torque-field}
\end{align}
\end{subequations}
are the scalar magnitudes of the interaction force and torque fields, as obtained from \cref{eq particle-interactions}.

With the collective force and torque given by \cref{eq collective-psi2}, \cref{eq psi1} is an integro-differential equation for $\Psi_1$ that involves $\Psi_2$. Therefore, \cref{eq psi1} is the first equation in the BBGKY hierarchy. The simplest approximation to truncate this hierarchy is Boltzmann's molecular-chaos approximation, whereby pair correlations are ignored and $\Psi_2$ is expressed in terms of $\Psi_1$. In active matter, however, pair correlations are crucial to capture the density-induced slowdown of particle motion, which is the original mechanism for motility-induced phase separation \cite{Bialke2013,Speck2020}. We thus go beyond the molecular-chaos approximation and account for pair correlations. To do this, one should derive an integro-differential equation for $\Psi_2$, which would involve $\Psi_3$, and truncate the hierarchy at that level to obtain a closed expression for $\Psi_2$ \cite{Hartel2018}. Here, however, instead of obtaining $\Psi_2$ from its own equation in the hierarchy, we directly measure the pair correlation function in experiments (\cref{Fig gr,parameters}) and use it as an input for the theory.

Specifically, to express \cref{eq psi1} as a closed equation for $\Psi_1$, we decompose $\Psi_2$ as
\begin{equation}
\Psi_2(\bm{r}_1,\theta_1,\bm{r}';t) = \rho(\bm{r}') g(\bm{r}' | \,\bm{r}_1,\theta_1;t) \Psi_1(\bm{r}_1,\theta_1;t).
\end{equation}
Here, $\Psi_2(\bm{r}_1,\theta_1,\bm{r}';t)$ is the density of particle pairs with one particle at position $\bm{r}_1$ with orientation $\theta_1$ and another particle at position $\bm{r}'$ with any orientation. Respectively, $\rho$ is the density field, and $g$ is the conventional dimensionless pair distribution function, so that $\rho(\bm{r}') g(\bm{r}' | \,\bm{r}_1,\theta_1)$ is the conditional density of particles at position $\bm{r}'$ given that another particle is at position $\bm{r}_1$ with orientation $\theta_1$. Finally, $\Psi_1(\bm{r}_1,\theta_1)$ is the density of particles at position $\bm{r}_1$ with orientation $\theta_1$. Next, we express $g$ in terms of the distance $|\bm{r}'-\bm{r}_1|$ between particles and the angle $\varphi$ formed between the vector $\bm{r}'-\bm{r}_1$ that joins both particle centers and the orientation vector $\hat{\bm{n}}_1$ of particle 1:
\begin{equation}
\hat{\bm{n}}_1\cdot \frac{\bm{r}'-\bm{r}_1}{|\bm{r}'-\bm{r}_1|} = \cos\varphi.
\end{equation}
Moreover, we assume that pair correlations depend only on the relative coordinates of the particle pair, i.e. their distance $|\bm{r}'-\bm{r}_1|$ and angle $\varphi$, and not on the coordinates $\bm{r}_1,\theta_1$ of the tagged particle 1. This assumption is verified in homogeneous states, and is therefore valid when we study the stability of the uniform isotropic state of the active Janus particle suspension. Finally, we focus on steady states, such that probability distributions are time-independent, and we drop the time variable hereafter. Mathematically, our assumptions read
\begin{equation}
g(\bm{r}' | \,\bm{r}_1,\theta_1) = g(|\bm{r}'-\bm{r}_1|,\varphi |\, \bm{r}_1,\theta_1) = g(|\bm{r}'-\bm{r}_1|,\varphi).
\end{equation}
With this decomposition of the two-particle distribution function $\Psi_2$, the collective force and torque (\cref{eq collective-psi2}) are expressed as
\begin{subequations} \label{eq collective-g}
\begin{align}
&\begin{multlined}
\bm{F}_{\text{int}}(\bm{r}_1,\theta_1) =  -\Psi_1(\bm{r}_1,\theta_1)\\
\times \int \dd^2\bm{r}' \; F(\left| \bm{r}'-\bm{r}_1\right|)\, \frac{\bm{r}'-\bm{r}_1}{\left| \bm{r}'-\bm{r}_1\right|} \,\rho(\bm{r}') \, g(|\bm{r}'-\bm{r}_1|,\varphi),
\end{multlined}
\\
&\begin{multlined}
\bm{\Gamma}_{\text{int}}(\bm{r}_1,\theta_1) =  \Psi_1(\bm{r}_1,\theta_1)\\
\times \int \dd^2\bm{r}' \; \Gamma(\left| \bm{r}'-\bm{r}_1\right|)\, \hat{\bm{n}}_1 \times \frac{\bm{r}'-\bm{r}_1}{\left| \bm{r}'-\bm{r}_1\right|}\, \rho(\bm{r}')\, g(|\bm{r}'-\bm{r}_1|,\varphi).
\end{multlined}
\end{align}
\end{subequations}

\subsubsection{Gradient expansion. Density-dependent collective force and torque} \label[SI section]{gradient-expansion}

Via \cref{eq collective-g}, the collective force and torque depend non-locally on the density field $\rho(\bm{r}')$. To derive local hydrodynamic equations, we perform a gradient expansion on the density field:
\begin{equation} \label{eq gradient-expansion}
\rho(\bm{r}')\approx \rho(\bm{r}_1) + \bm{\nabla}\rho(\bm{r}_1) \cdot(\bm{r}'-\bm{r}_1).
\end{equation}
Based on this expansion, we first obtain the zeroth-order contribution in the density gradient to the collective force and torque. Introducing \cref{eq gradient-expansion} into \cref{eq collective-g}, and changing the integration variable to the relative position $\bm{r}\equiv \bm{r}'-\bm{r}_1$, the zeroth-order term gives
\begin{subequations}
\begin{align}
&\begin{multlined}
\bm{F}^{(0)}_{\text{int}}(\bm{r}_1,\theta_1) = -\Psi_1(\bm{r}_1,\theta_1) \rho(\bm{r}_1)\\
\times \int_0^\infty \dd r \; r\, F(r) \int_0^{2\pi} \dd\varphi\; g(r,\varphi) \,\hat{\bm{r}}(\varphi),
\end{multlined}\\
&\begin{multlined}
\bm{\Gamma}^{(0)}_{\text{int}}(\bm{r}_1,\theta_1) = \Psi_1(\bm{r}_1,\theta_1) \rho(\bm{r}_1)\\
\times \int_0^\infty \dd r \; r\, \Gamma(r) \int_0^{2\pi} \dd\varphi\; g(r,\varphi) \sin\varphi \,\hat{\bm{z}},
\end{multlined}
\end{align}
\end{subequations}
where we have used that $\hat{\bm{n}}_1\times \hat{\bm{r}} = \sin \varphi \,\hat{\bm{z}}$. For passive fluids with isotropic interactions, the pair distribution function is isotropic, $g(r,\varphi) = g(r)$. However, for active fluids, even with isotropic interactions, $g(r,\varphi)$ is anisotropic. This anisotropy is due to particle self-propulsion, which determines a direction and implies that other particles are more likely to be found in front than behind the tagged particle \cite{Bialke2013}. In the absence of any chirality, $g(r,\varphi)$ is symmetric around the particle self-propulsion axis, i.e. around $\varphi=0$. Therefore, $g(r,\varphi)$ is an even function of $\varphi$. As a consequence, the component of the collective force $\bm{F}_{\text{int}}^{(0)}$ along the self-propulsion direction, $\bm{F}_{\text{int}}^{(0)}\cdot \hat{\bm{n}}_1$, is non-zero because $\hat{\bm{n}}_1\cdot \hat{\bm{r}} = \cos\varphi$, and hence $g(r,\varphi)\cos\varphi$ is an even function of $\varphi$. In contrast, the component of the collective force perpendicular to the self-propulsion direction vanishes because $g(r,\varphi) \sin\varphi$ is an odd function of $\varphi$. For the same reason, the zeroth-order collective torque $\bm{\Gamma}_{\text{int}}^{(0)}$ vanishes. Altogether:
\begin{subequations}
\begin{align}
&\begin{multlined}
\bm{F}^{(0)}_{\text{int}}(\bm{r}_1,\theta_1) = -\Psi_1(\bm{r}_1,\theta_1) \rho(\bm{r}_1) \hat{\bm{n}}_1\\
\times \int_0^\infty \dd r \; r\, F(r) \int_0^{2\pi} \dd\varphi\; g(r,\varphi) \cos\varphi,
\end{multlined}
\\
&\bm{\Gamma}^{(0)}_{\text{int}}(\bm{r}_1,\theta_1) = 0.
\end{align}
\end{subequations}
$\bm{\Gamma}^{(0)}_{\text{int}}$ corresponds to the collective torque experienced by a probe particle in a uniform-density background. Because interaction torques make particles orient toward the location of other particles (\cref{eq interaction-torque}), particles in a uniform density field have no preferred direction to orient toward. Therefore, the zeroth-order contribution to the collective torque vanishes.

Next, we obtain the first-order contribution in the density gradient to the collective force and torque. Using the same symmetry arguments as for the zeroth-order contribution, we obtain
\begin{subequations}
\begin{align}
&\begin{multlined}
\bm{F}^{(1)}_{\text{int},\parallel}(\bm{r}_1,\theta_1) = -\Psi_1(\bm{r}_1,\theta_1) \bm{\nabla}_\parallel\rho(\bm{r}_1)\\
\times \int_0^\infty \dd r \; r^2\, F(r) \int_0^{2\pi} \dd\varphi\; g(r,\varphi)\cos^2\varphi,
\end{multlined}
\\
&\begin{multlined}
\bm{F}^{(1)}_{\text{int},\perp}(\bm{r}_1,\theta_1) = -\Psi_1(\bm{r}_1,\theta_1) \bm{\nabla}_\perp\rho(\bm{r}_1)\\
\times \int_0^\infty \dd r \; r^2\, F(r) \int_0^{2\pi} \dd\varphi\; g(r,\varphi)\sin^2\varphi,
\end{multlined}
\end{align}
\end{subequations}
for the force components parallel and perpendicular to the particle orientation $\hat{\bm{n}}_1$. The integrals in the parallel and perpendicular components are different. However, for typical forms of $g(r,\varphi)$, both integrals take similar values. Thus, to simplify the calculation, we ignore the small difference between these two integrals, replacing them by the average of both integrals. With this approximation, $\bm{F}^{(1)}_{\text{int}}$ is directly proportional to the density gradient. Respectively, for the collective torque, we obtain
\begin{multline}
\bm{\Gamma}^{(1)}_{\text{int}}(\bm{r}_1,\theta_1) = \Psi_1(\bm{r}_1,\theta_1)\, \hat{\bm{n}}_1\times \bm{\nabla}\rho(\bm{r}_1)\\
\times \int_0^\infty \dd r \; r^2\, \Gamma(r) \int_0^{2\pi} \dd\varphi\; g(r,\varphi) \sin^2\varphi.
\end{multline}

Gathering the results for the zeroth- and first-order contributions, we finally have
\begin{subequations} \label{eq collective-force-torque-final}
\begin{align}
\bm{F}_{\text{int}}(\bm{r}_1,\theta_1) &\approx -\Psi_1(\bm{r}_1,\theta_1) \left[\zeta_0 \rho(\bm{r}_1)\hat{\bm{n}}_1 + \zeta_1 \bm{\nabla}\rho(\bm{r}_1)\right],\label{eq collective-force-final}\\
\bm{\Gamma}_{\text{int}}(\bm{r}_1,\theta_1) &= \Psi_1(\bm{r}_1,\theta_1)\, \tau_1\, \hat{\bm{n}}_1\times \bm{\nabla}\rho(\bm{r}_1) \label{eq collective-torque-final}.
\end{align}
\end{subequations}
Here, we have defined the coefficients
\begin{subequations} \label{eq coefficients}
\begin{align}
\zeta_0 &\equiv \int_0^\infty \dd r \; r\, F(r) \int_0^{2\pi} \dd\varphi\; g(r,\varphi) \cos\varphi, \label{eq zeta0}\\
\zeta_1 &\equiv \frac{1}{2} \int_0^\infty \dd r \; r^2\, F(r) \int_0^{2\pi} \dd\varphi\; g(r,\varphi), \label{eq zeta1}\\
\tau_1 &\equiv \int_0^\infty \dd r \; r^2\, \Gamma(r) \int_0^{2\pi} \dd\varphi\; g(r,\varphi) \sin^2\varphi, \label{eq tau1}
\end{align}
\end{subequations}
with $F(r)$ and $\Gamma(r)$ given in \cref{eq force-torque-fields-expressions}. These coefficients capture the effects of interparticle interactions at the coarse-grained level, taking into account the anisotropy of the pair distribution function in systems of self-propelled particles. Specifically, the collective force and torque terms characterized by each of these coefficients (see \cref{eq collective-force-torque-final}) capture the following effects:
\begin{itemize}
\item \emph{Repulsion-induced slowdown}: When the net central force between particles is repulsive ($F(r)>0$), the term with $\zeta_0>0$ gives a force that opposes particle self-propulsion due to the higher probability of finding other particles in front than behind a given particle. The higher the particle density, the higher the opposing force. This opposing force causes the density-dependent slowdown of particle motion that is responsible for standard motility-induced phase separation\cite{Cates2015,Speck2020}.
\item \emph{Repulsion-induced diffusion}: For net repulsive interparticle forces ($F(r)>0$), the term with $\zeta_1>0$ gives a force opposed to density gradients, tending to homogenize particle concentration like a diffusive flux.
\item \emph{Torque toward denser regions}: When the tail dipole has a larger magnitude than the head dipole, $d_{\text{t}}^2>d_{\text{h}}^2$, we have $\Gamma(r)>0$ (see \cref{eq torque-field}), and the term with $\tau_1$ gives a torque that tends to align particle orientation $\hat{\bm{n}}_1$ with the density gradient, thus reorienting particle motion toward higher-density regions. Therefore, the collective torque \cref{eq collective-torque-final} encodes the effect that is responsible for the new mechanism of motility-induced phase separation that we unveil here.
\end{itemize}
Whereas the repulsion-induced slowdown was obtained in previous studies\cite{Bialke2013,Speck2020}, here we account both for repulsion-induced diffusion and, crucially, for torque toward denser regions by performing the gradient expansion up to first order in the density gradient.

To end this subsection, we introduce the final expressions for the collective force and torque (\cref{eq collective-force-torque-final}) into the Smoluchowski equation (\cref{eq psi1}) for the one-particle distribution function $\Psi_1$. We obtain
\begin{multline} \label{eq psi1-final}
\partial_t \Psi_1 = -\bm{\nabla}\cdot\left\{\left[\left( v_0 - \frac{\zeta_0}{\xi_{\text{t}}}\rho \right)\hat{\bm{n}} - \frac{\zeta_1}{\xi_{\text{t}}} \bm{\nabla}\rho - D_{\text{t}}\,\bm{\nabla}\right] \Psi_1\right\} \\
- \partial_\theta \left\{ \left[ \frac{\tau_1}{\xi_{\text{r}}} \hat{\bm{z}}\cdot \left(\hat{\bm{n}}\times \bm{\nabla}\rho\right) - D_{\text{r}} \,\partial_\theta\right] \Psi_1\right\}.
\end{multline}
Here, we have dropped the subscript $1$ to indicate the tagged particle. We do that hereafter.

\subsubsection{Torque non-reciprocity produces particle alignment toward denser regions} \label[SI section]{non-reciprocal}

Before coarse-graining the model further, we note that the non-reciprocal character of the interaction torques (\cref{eq interaction-torque}), i.e. the fact that $\bm{\Gamma}_{ij} \neq -\bm{\Gamma}_{ji}$, is important for the reorientation of particles toward denser regions. Torques result from the electrostatic interactions between the dipoles in each hemisphere of the Janus particles. Therefore, the torque between particles $i$ and $j$ must be built based on three vectors: the distance vector $\bm{r}_{ij}$, and the self-propulsion direction of each particle, $\hat{\bm{n}}_i$ and $\hat{\bm{n}}_j$, which indicates the direction in which the dipoles are displaced from the particle's center of mass. Because the dipoles are perpendicular to the interparticle distance, they only exert in-plane forces (\cref{eq dipolar-force}). Hence, particles will rotate in their plane of motion, implying that the interaction torque points out of plane (along $\hat{\bm{z}}$). To build an out-of-plane vector from the three in-plane vectors $\bm{r}_{ij}$, $\hat{\bm{n}}_i$, and $\hat{\bm{n}}_j$, we must take vector products between them. Thus, the interaction torque could potentially be either $\bm{\Gamma}_{ij} \propto \hat{\bm{n}}_i \times \hat{\bm{r}}_{ij}$, $\bm{\Gamma}_{ij} \propto \hat{\bm{n}}_j \times \hat{\bm{r}}_{ij}$, or $\bm{\Gamma}_{ij} \propto \hat{\bm{n}}_i \times \hat{\bm{n}}_{j}$.

The two first torques are non-reciprocal ($\bm{\Gamma}_{ij} \neq -\bm{\Gamma}_{ji}$); they orient one particle toward the location of the other particle. The third torque is reciprocal ($\bm{\Gamma}_{ij} = -\bm{\Gamma}_{ji}$); it tends to align the particles. Because this reciprocal torque does not depend on the direction in which the other particle is located, it cannot orient one particle toward the location of the other. Hence, reciprocal torques based on the electrostatic interactions in our system cannot reorient particles toward denser regions, and therefore they cannot lead to the torque-based phase separation that we report here. In other words, for reciprocal torques $\bm{\Gamma}_{ij} \propto \hat{\bm{n}}_i \times \hat{\bm{n}}_{j}$, the collective torque $\bm{\Gamma}_{\text{int}}$ would vanish \cite{Jayaram2020,Grossmann2020}.

\subsubsection{Moment hierarchy and hydrodynamic equations} \label[SI section]{hydrodynamic-equations}

To complete the coarse-graining of the microscopic model, we define continuum fields as the angular moments of the one-particle distribution function $\Psi_1$. For example, the zeroth moment corresponds to the density field $\rho(\bm{r},t)$, the first moment corresponds to the polarization density $\bm{p}(\bm{r},t)$, and the second moment is related to the nematic order-parameter tensor density $\bm{Q}(\bm{r},t)$:
\begin{subequations}
\begin{align}
\rho(\bm{r},t) &=\int \Psi_1(\bm{r},\theta,t)\, \dd\theta,\\
p_\alpha(\bm{r},t) &=\int n_\alpha\, \Psi_1(\bm{r},\theta,t)\, \dd\theta,\\
Q_{\alpha\beta}(\bm{r},t) &=\int \left[n_\alpha n_\beta-\frac{1}{2}\delta_{\alpha\beta}\right] \Psi_1(\bm{r},\theta,t)\, \dd\theta.
\end{align}
\end{subequations}
In general, the $k^{\text{th}}$ moment of $\Psi_1$ with respect to the particle orientation $\hat{\bm{n}}$ corresponds to a $k^{\text{th}}$-rank orientational tensor field. We can then obtain hydrodynamic equations for each of these fields by taking the corresponding moment of the Smoluchowski equation \cref{eq psi1-final}. Proceeding in this way, we obtain the following equations for the density and the polarity fields:
\begin{subequations} \label{eq hydrodynamic-equations-before-truncation}
\begin{align}
&\partial_t \rho= -\bm{\nabla}\cdot\left[\left( v_0 - \frac{\zeta_0}{\xi_{\text{t}}}\rho \right) \bm{p}\right] + \bm{\nabla}\cdot\left( \frac{\zeta_1}{\xi_{\text{t}}}\rho \bm{\nabla}\rho \right) + D_{\text{t}}\nabla^2\rho, \label{eq density}\\
&\begin{multlined} \label{eq polarity}
\partial_t \bm{p} = -\bm{\nabla}\cdot\left[\left( v_0 - \frac{\zeta_0}{\xi_{\text{t}}}\rho \right) \bm{Q} \right] - \frac{1}{2}\bm{\nabla}\left[ \left( v_0 - \frac{\zeta_0}{\xi_{\text{t}}}\rho \right) \rho \right]\\
 + \bm{\nabla}\cdot\left(\frac{\zeta_1}{\xi_{\text{t}}} \bm{p} \bm{\nabla}\rho\right) + D_{\text{t}}\nabla^2\bm{p} \\
 -\frac{\tau_1}{\xi_{\text{r}}}\left( \bm{Q}\cdot\bm{\nabla}\rho - \frac{1}{2}\rho\bm{\nabla}\rho \right) - D_{\text{r}}\, \bm{p}.
\end{multlined}
\end{align}
\end{subequations}
In general, the equation for the $k^{\text{th}}$ moment involves the $k+1^{\text{th}}$ moment, giving rise to a hierarchy of hydrodynamic equations.

To close this hierarchy, we eliminate the nematic tensor $\bm{Q}$ in terms of the density and polarity fields. To this end, we obtain the equation for $\bm{Q}$:
\begin{multline}
\partial_t \bm{Q} = \frac{v_0}{2}\left[ -2\bm{\nabla}\cdot\bm{T} - \bm{I}_3\cdot\bm{\nabla}\rho + \bm{I}\bm{\nabla}\cdot\bm{p} \right] \\
+ \frac{\zeta_1}{\xi_{\text{t}}} \bm{\nabla}\cdot\left[(\bm{\nabla}\rho)\bm{Q}\right] + D_{\text{t}}\nabla^2\bm{Q} - D_{\text{r}}\bm{Q}\\
+ \frac{\tau_1}{\xi_{\text{r}}} \left[(\bm{\nabla}\rho)\bm{p} + \bm{p}(\bm{\nabla}\rho) - 4\bm{T}\cdot\bm{\nabla}\rho - 2 \rho\,\bm{I}_3\cdot\bm{\nabla}\rho - \bm{I}\bm{p}\cdot\bm{\nabla}\rho\right],
\end{multline}
where $\bm{I}_3$ is the third-rank identity tensor, and
\begin{equation}
T_{\alpha\beta\gamma}(\bm{r},t)=\int \left[n_\alpha n_\beta n_\gamma - \frac{1}{2}\delta_{\alpha\beta\gamma}\right]\Psi_1(\bm{r},\theta,t)\,\dd\theta
\end{equation}
is the traceless third-rank tensor corresponding to the third-order orientational moment of $\Psi_1$. A standard closure approximation then corresponds to setting $\bm{T} = \bm{0}$, and eliminating $\bm{Q}$ by imposing $\partial_t \bm{Q}\approx \bm{0}$. This approximation is based on the fact that $\bm{Q}$ relaxes faster than $\bm{p}$ and $\rho$. Moreover, working in the hydrodynamic limit, we neglect terms with higher-order gradients of $\bm{Q}$ in front of the relaxation term $2D_{\text{r}}\bm{Q}$. This way, $\bm{Q}$ can be expressed in terms of the lower-order moments $\rho$ and $\bm{p}$ as
\begin{multline} \label{eq Q}
D_{\text{r}}\bm{Q}\approx \frac{v_0}{2}\left[\bm{I}\bm{\nabla}\cdot\bm{p} - \bm{I}_3\cdot\bm{\nabla}\rho\right] \\
+ \frac{\tau_1}{\xi_{\text{r}}} \left[(\bm{\nabla}\rho)\bm{p} + \bm{p}(\bm{\nabla}\rho) - 2 \rho\,\bm{I}_3\cdot\bm{\nabla}\rho - \bm{I}\bm{p}\cdot\bm{\nabla}\rho\right],
\end{multline}
and hence eliminated from \cref{eq polarity}. In practice, however, all the terms coming from $\bm{Q}$ are of higher order in gradients than other terms in \cref{eq polarity}. Therefore, these $\bm{Q}$ terms can be neglected in the hydrodynamic limit, which we focus on to predict the phase diagram of the system. Hence, for our practical purposes, we set $\bm{Q}=\bm{0}$. We then recast the hydrodynamic equations \cref{eq hydrodynamic-equations-before-truncation} as
\begin{subequations} \label{eq hydrodynamic-equations}
\begin{align}
&\partial_t \rho = -\bm{\nabla}\cdot\left(v[\rho] \bm{p}\right) + \bm{\nabla}\cdot\left((D_{\text{t}} + D_{\text{rep}}[\rho]) \bm{\nabla}\rho \right), \label{eq density-new}\\
&\begin{multlined}
\partial_t \bm{p} = - D_{\text{r}}\, \bm{p} + D_{\text{t}} \nabla^2\bm{p} - \frac{1}{2}\bm{\nabla}\left( v[\rho] \rho \right) + \frac{1}{2} v_{\text{tor}}[\rho] \bm{\nabla}\rho \\
+ \frac{D_{\text{rep}}[\rho]}{\rho} \bm{\nabla}\cdot\left( \bm{p} \bm{\nabla}\rho\right). \label{eq polarity-new}
\end{multlined}
\end{align}
\end{subequations}
These equations explicitly showcase the effects of the different collective force and torque contributions at the hydrodynamic level. Specifically, the three effects listed in the previous subsection are encoded into three functionals,
\begin{subequations} \label{eq interaction-functionals}
\begin{align}
v[\rho(\bm{r})] &= v_0 - \zeta_0 \rho(\bm{r})/\xi_{\text{t}}, \label{eq local-speed}\\
D_{\text{rep}}[\rho(\bm{r})] &= \zeta_1 \rho(\bm{r})/\xi_{\text{t}}, \label{eq interaction-diffusion}\\
v_{\text{tor}}[\rho(\bm{r})] &= \tau_1 \rho(\bm{r})/\xi_{\text{r}}, \label{eq interaction-speed}
\end{align}
\end{subequations}
with the following interpretations:
\begin{itemize}
\item The repulsion-induced slowdown is apparent in the density-dependent particle speed $v[\rho]$ (\cref{eq local-speed}), which decreases with increasing density
\item The repulsion-induced diffusion shows up as additional density-dependent diffusion coefficient $D_{\text{rep}}[\rho]$ (\cref{eq interaction-diffusion})
\item The torque toward denser regions manifests as an additional density-dependent speed $v_{\text{tor}}[\rho]$ (\cref{eq interaction-speed}) for the polarity field, thus contributing to aligning polarity with the density gradient.
\end{itemize}
The coefficients $\zeta_0$, $\zeta_1$, and $\tau_1$ that characterize each of these effects are given in \cref{eq coefficients}.

\Cref{eq hydrodynamic-equations} are hydrodynamic equations for the coupled density and polarity fields of our suspension of interacting active particles. These equations are the final outcome of our derivation. Via the coarse-graining procedure presented in this section, we have obtained all the coefficients in \cref{eq hydrodynamic-equations}, which characterize the large-scale collective behavior of our system, in terms of the microscopic parameters of our active particles.

\subsection{Prediction of the phase diagram} \label[SI section]{phase-diagram}

In this section, we employ the hydrodynamic description obtained in \cref{coarse-graining} to predict the phase diagram of our system (\cref{Fig diagram}). We obtain the phase diagram in terms of the microscopic parameters characterizing the self-propulsion and electrostatic interactions of our active particles, which we measure in experiments.

\subsubsection{Stability of the uniform state. Spinodal lines} \label[SI section]{spinodal}

The simplest steady state of our active particle suspension is an active gas with a homogeneous particle density $\rho=\rho_0$ and no polarity, $\bm{p}=\bm{0}$. In this section, we analyze the stability of this uniform and isotropic state. Thereby, we obtain the spinodal region of the phase diagram, where the uniform state is linearly unstable to phase separation.

To analyze the stability of the uniform state to slow, long-wavelength perturbations, we focus on the density field, which is the only slow variable of the problem. Respectively, the polarity field $\bm{p}$ relaxes over a finite time scale $D_{\text{r}}^{-1}$, becoming adiabatically enslaved to the density field. From \cref{eq polarity-new}, and to lowest order in gradients, the polarity tends to
\begin{equation}
\bm{p}_{\text{ad}} = \frac{1}{2D_{\text{r}}} \left( v_{\text{tor}} [\rho] \bm{\nabla}\rho - \bm{\nabla}(v[\rho] \rho)\right).
\end{equation}
Introducing this expression into \cref{eq density-new}, we obtain a closed equation for the density field, which can be written as
\begin{equation} \label{eq adiabatic-density}
\partial_t \rho = -\bm{\nabla}\cdot\bm{J};\qquad \bm{J} = -\mathcal{D}[\rho] \bm{\nabla}\rho,
\end{equation}
where
\begin{equation} \label{eq collective-diffusivity}
\mathcal{D}[\rho] = D_{\text{t}} + D_{\text{rep}}[\rho] + \frac{v[\rho]}{2D_{\text{r}}} \left( v[\rho] + v'[\rho]\rho - v_{\text{tor}}[\rho] \right)
\end{equation}
is a collective diffusivity functional.

Based on \cref{eq adiabatic-density}, density perturbations $\delta\rho(\bm{r},t) = \rho(\bm{r},t) - \rho_0$ around the uniform density $\rho_0$ evolve as
\begin{equation}
\partial_t \delta\rho = \mathcal{D}(\rho_0) \nabla^2 \delta\rho.
\end{equation}
Thus, the uniform state experiences a spinodal instability for $\mathcal{D}(\rho_0)<0$. In the absence of interaction torques ($v_{\text{tor}}[\rho] = 0$), the collective diffusivity $\mathcal{D}(\rho_0)$ can turn negative due to the repulsion-induced slowdown, which implies $v'[\rho]<0$. This effect is the original mechanism for motility-induced phase separation in systems of repulsive self-propelled particles \cite{Cates2015}. Here, via \cref{eq collective-diffusivity}, we show that, even in the absence of repulsion-induced slowdown ($v'[\rho]=0$), interaction torques (with $v'_{\text{tor}}>0$) can induce motility-induced phase separation.

Including both repulsion and torques, the spinodal lines are determined by the condition $\mathcal{D}(\rho_0) = 0$. In the plane of bare self-propulsion and particle density $(v_0,\rho_0)$, the spinodal lines are given by
\begin{multline} \label{eq spinodal}
v_0^{\text{sp}} = \frac{1}{2}\left[ (v'_{\text{tor}} - 3 v'_{\text{rep}})\rho_0 \phantom{\pm \sqrt{ (v'_{\text{tor}} - v'_{\text{rep}})^2\rho_0^2 - 8 D_{\text{r}} (D_{\text{t}} + D'_{\text{rep}} \rho_0)}} \right.\\
\left.\pm \sqrt{ (v'_{\text{tor}} - v'_{\text{rep}})^2\rho_0^2 - 8 D_{\text{r}} (D_{\text{t}} + D'_{\text{rep}} \rho_0)}\,\right],
\end{multline}
which are shown in blue in \cref{Fig diagram,Fig slowdown-only,Fig torques-only}. Here, based on \cref{eq interaction-functionals}, we have defined the density-independent parameters
\begin{subequations} \label{eq slopes}
\begin{align}
&v'_{\text{rep}} = -\zeta_0/\xi_{\text{t}} <0,\\
&D'_{\text{rep}} = \zeta_1/\xi_{\text{t}} >0,\\
&v'_{\text{tor}} = \tau_1/\xi_{\text{r}} >0.
\end{align}
\end{subequations}
In \cref{t parameters}, we provide estimates for the values of these parameters in our experiments.

\subsubsection{Phase coexistence. Binodal lines} \label[SI section]{binodals}

The spinodal instability presented in \cref{spinodal} leads to phase separation. In this subsection, we build an effective thermodynamic description to predict the densities of the coexisting phases, namely the binodal lines of the phase diagram.

To obtain the binodals, we consider the closed long-time dynamics of the density field given by \cref{eq adiabatic-density}. Following previous work \cite{Cates2015,Stenhammar2013,Wittkowski2014,Speck2014,Speck2015}, we recognize this equation as an effective Cahn-Hilliard equation, which implies the existence of an effective free energy governing phase separation. To obtain this effective free energy, we express the density current in \cref{eq adiabatic-density} as deriving from an effective chemical potential $\mu[\rho]$:
\begin{equation} \label{eq CH-flux}
\bm{J} = -M[\rho]\bm{\nabla} \mu[\rho] = -M[\rho]\mu'[\rho]\bm{\nabla}\rho,
\end{equation}
where $M[\rho] = \beta D[\rho]\rho$ is the collective mobility associated with the long-time diffusivity $D[\rho]$, with $\beta = 1/(k_BT)$. To obtain this diffusivity, we use that our hydrodynamic description corresponds to a system of self-propelled particles with density-dependent self-propulsion speed $v[\rho]$ and translational diffusion coefficient $D_{\text{t}} + D_{\text{rep}}[\rho]$ (see \cref{eq density-new}). These particles undergo a random walk with persistence time $D_{\text{r}}^{-1}$ and with a long-time diffusion coefficient given by
\begin{equation}
D[\rho] = D_t + D_{\text{rep}}' \rho + \frac{v^2[\rho]}{2D_r}.
\end{equation}
Using this expression to obtain the mobility $M[\rho]$, and comparing \cref{eq CH-flux} with \cref{eq adiabatic-density}, we obtain
\begin{equation}
\mu'(\rho) = \frac{\mathcal{D}(\rho)}{M(\rho)} =  \frac{k_BT}{\rho} \left[ 1 + \frac{\frac{v(\rho)}{2D_{\text{r}}} [ v'_{\text{rep}} - v'_{\text{tor}}] \rho}{D_{\text{t}} + D'_{\text{rep}} \rho + \frac{v^2(\rho)}{2D_{\text{r}}}}\right].
\end{equation}
Integrating over $\rho$, we obtain the chemical potential
\begin{multline} \label{eq chemical-potential}
\beta\mu(\rho) = \ln \rho + \frac{1}{2} \left[ 1 - \frac{v'_{\text{tor}}}{v'_{\text{rep}}} \right] \ln D(\rho) \\
+ \frac{D_{\text{r}} D'_{\text{rep}} [1- v'_{\text{tor}}/v'_{\text{rep}}]}{\sqrt{ D_{\text{r}}^2 (D'_{\text{rep}})^2 + 2v'_{\text{rep}} D_{\text{r}} (v_0 D'_{\text{rep}} - v'_{\text{rep}} D_{\text{t}})}} \\
 \times \arctanh \left( \frac{D_{\text{r}} D'_{\text{rep}} + v'_{\text{rep}} v(\rho)}{\sqrt{ D_{\text{r}}^2 (D'_{\text{rep}})^2 + 2v'_{\text{rep}} D_{\text{r}} (v_0 D'_{\text{rep}} - v'_{\text{rep}} D_{\text{t}})}}\right)
\end{multline}
up to an irrelevant constant term. Assuming that the free energy density $f(\rho)$ is a local function of the density, it obeys $f'(\rho) = \mu(\rho)$. Integrating this equation, we obtain
\begin{multline} \label{eq free-energy}
\beta f(\rho) = \rho (\ln \rho - 1) \\
+ \frac{1}{2}\left[ 1 - \frac{v'_{\text{tor}}}{v'_{\text{rep}}}\right] \left[ \rho + \frac{2D_{\text{r}} D'_{\text{rep}} + v_0 v'_{\text{rep}}}{(v'_{\text{rep}})^2} \right] \ln D(\rho)\\
+ \frac{D_{\text{r}} D'_{\text{rep}} [1- v'_{\text{tor}}/v'_{\text{rep}}] \left[\rho + \frac{2D_{\text{r}} D'_{\text{rep}} + 3 v_0 v'_{\text{rep}}}{(v'_{\text{rep}})^2} - \frac{2D_{\text{t}}}{D'_{\text{rep}}}\right]}{\sqrt{ D_{\text{r}}^2 (D'_{\text{rep}})^2 + 2v'_{\text{rep}} D_{\text{r}} (v_0 D'_{\text{rep}} - v'_{\text{rep}} D_{\text{t}})}} \\
\times \arctanh \left( \frac{D_{\text{r}} D'_{\text{rep}} + v'_{\text{rep}} v(\rho)}{\sqrt{ D_{\text{r}}^2 (D'_{\text{rep}})^2 + 2v'_{\text{rep}} D_{\text{r}} (v_0 D'_{\text{rep}} - v'_{\text{rep}} D_{\text{t}})}}\right)
\end{multline}
up to irrelevant terms linear in the density $\rho$. The thermodynamic pressure can then be obtained as $p(\rho) = \mu(\rho) \rho - f(\rho)$, giving
\begin{multline} \label{eq pressure}
\beta p(\rho) = \rho + \left[ 1 - \frac{v'_{\text{tor}}}{v'_{\text{rep}}}\right] \left[ \rho - \frac{1}{2}\frac{2D_{\text{r}} D'_{\text{rep}} + v_0 v'_{\text{rep}}}{(v'_{\text{rep}})^2}\right] \ln D(\rho)\\
- \frac{D_{\text{r}} D'_{\text{rep}} [1- v'_{\text{tor}}/v'_{\text{rep}}] \left[\frac{2D_{\text{r}} D'_{\text{rep}} + 3 v_0 v'_{\text{rep}}}{(v'_{\text{rep}})^2} - \frac{2D_{\text{t}}}{D'_{\text{rep}}}\right]}{\sqrt{ D_{\text{r}}^2 (D'_{\text{rep}})^2 + 2v'_{\text{rep}} D_{\text{r}} (v_0 D'_{\text{rep}} - v'_{\text{rep}} D_{\text{t}})}} \\
\times \arctanh \left( \frac{D_{\text{r}} D'_{\text{rep}} + v'_{\text{rep}} v(\rho)}{\sqrt{ D_{\text{r}}^2 (D'_{\text{rep}})^2 + 2v'_{\text{rep}} D_{\text{r}} (v_0 D'_{\text{rep}} - v'_{\text{rep}} D_{\text{t}})}}\right).
\end{multline}
up to an irrelevant constant. These three effective thermodynamic functions, namely the chemical potential $\mu(\rho)$, the free energy density $f(\rho)$, and the pressure $p(\rho)$ are plotted in \cref{Fig thermodynamics}.

Leveraging this effective thermodynamic picture, we obtain the binodal lines by building a common-tangent construction on the effective free energy $f(\rho)$ (\cref{fig free-energy}). Given that $f(\rho) = \mu(\rho) \rho - p(\rho)$, the common-tangent construction corresponds to requiring coexisting phases to have equal chemical potential and pressure,
\begin{equation} \label{eq binodal}
\mu(\rho_g) = \mu(\rho_l),\qquad p(\rho_g) = p(\rho_l),
\end{equation}
where $\rho_g$ and $\rho_l$ are the densities of the coexisting (gas and liquid) phases (\cref{fig chemical-potential,fig pressure}). Alternatively, the common-tangent construction also amounts to performing the Maxwell construction on the curve $p(1/\rho)$ (\cref{fig pressure-Maxwell}). In practice, we solve \cref{eq binodal} numerically and obtain the binodal lines shown in red in \cref{Fig diagram,Fig slowdown-only}.

Using the thermodynamic framework presented above, we obtain the spinodal lines from the inflection points of the effective free energy, given by condition $f''(\rho) = 0$. The spinodal lines obtained in this way coincide with the results of the linear stability analysis in \cref{spinodal}, confirming the consistency of our thermodynamic approach.

Finally, although qualitatively correct, the binodals obtained from the common-tangent construction above have been shown to fail to accurately reproduce the coexisting densities measured in simulations of active Brownian particles \cite{Cates2015,Speck2020,Solon2015,Paliwal2018,Solon2018,Solon2018b}. In recent work, Solon et al. have developed a theory to predict quantitatively-accurate binodals for systems of active Brownian particles \cite{Solon2018,Solon2018b}. In that framework, the non-equilibrium corrections to the common-tangent construction stem from the fact that the effective chemical potential is a non-local functional of the density, including density-gradient terms that describe interfaces \cite{Cates2015,Wittkowski2014,Solon2018,Solon2018b}. While it is in principle possible to obtain these terms by means of a gradient expansion of \cref{eq polarity-new,eq collective-g} to order higher than we did in \cref{gradient-expansion}, this calculation falls beyond the scope of our work.

\subsection{Parameter estimates} \label[SI section]{parameters}

In this section, we provide experimental estimates for the parameters of the model, which we list in \cref{t parameters}. In the first part of \cref{t parameters}, we collect estimates for parameters that are measured quantities. In the second part of \cref{t parameters}, we collect the formulae to derive other parameter values from the measured parameters as explained below. In the third part of \cref{t parameters}, we collect the formulae and values of interaction parameters obtained numerically using the experimentally measured pair distribution function (\cref{Fig gr}) as explained below.

\begin{table*}[tb]
\begin{center}
\begin{tabular}{lc}

Description&Estimate\\\hline

Particle radius&$R\approx 1.5$ $\mu$m\\
Self-propulsion speed&$v_0\approx 10-40$ $\mu$m/s\\
Area fraction of particles&$\phi_0\approx 0.15$\\
Solvent viscosity&$\eta\approx 0.89$ mPa$\cdot$s\\
Rotational diffusion coefficient&$D_{\text{r}}\approx 0.15$ s$^{-1}$\\
Voltage amplitude&$V_0 = 10$ V\\
Electric field frequency&$\nu = 30$ kHz\\
Electrostatic screening length&$\lambda\approx 120$ $\mu$m\\
Dielectric permittivity of vacuum&$\epsilon_0\approx 8.85\cdot 10^{-12}$ C$^2$/(N m$^2$)\\
Relative dielectric permittivity of the solvent&$\epsilon_{\text{r}}\approx 78.5$\\
Real part of the head dipole factor&$\mathrm{Re}[K_{\text{h}}]\approx -0.38$\\
Imaginary part of the head dipole factor&$\mathrm{Im}[K_{\text{h}}]\approx 0.015$\\
Real part of the tail dipole factor&$\mathrm{Re}[K_{\text{t}}]\approx 0.28$\\
Imaginary part of the tail dipole factor&$\mathrm{Im}[K_{\text{t}}]\approx 0.75$\\\hline

Dipole shift distance&$\ell = 3R/8\approx 0.56$ $\mu$m\\
Particle number density&$\rho_0 = \phi_0/(\pi R^2) \approx 0.02$ $\mu$m$^{-2}$\\
Translational drag coefficient&$\xi_{\text{t}} = 6\pi \eta R\approx 25$ mPa$\cdot$s$\cdot\mu$m\\
Rotational drag coefficient&$\xi_{\text{r}} = 8\pi\eta R^3\approx 75$ mPa$\cdot$s$\cdot\mu$m$^3$\\
Translational diffusion coefficient&$D_{\text{t}} = k_BT/\xi_{\text{t}}\approx 0.16$ $\mu$m$^{-2}$/s\\
Electric field amplitude&$E_0 = V_0/\lambda \approx 83$ V/mm\\
Dielectric permittivity of the solvent&$\epsilon = \epsilon_{\text{r}}\epsilon_0\approx 6.95\cdot 10^{-10}$ C$^2$/(N m$^2$)\\\hline

Repulsion-induced slowdown force coefficient&$\zeta_0 \approx 12$ N$\cdot\mu$m$^2$, \cref{eq zeta0}\\
Repulsion-induced diffusion force coefficient&$\zeta_1 \approx 706$ N$\cdot\mu$m$^3$, \cref{eq zeta1}\\
Interaction torque coefficient&$\tau_1 \approx 300$ N$\cdot\mu$m$^4$, \cref{eq tau1}\\
Repulsion-induced slowdown effective speed&$v_{\text{rep}} = \zeta_0\rho_0/\xi_{\text{t}} \approx 10$ $\mu$m/s\\
Repulsion-induced effective diffusivity&$D_{\text{rep}} = \zeta_1 \rho_0/\xi_{\text{t}} \approx 595$ $\mu$m$^2$/s\\
Torque-induced effective speed&$v_{\text{tor}} = \tau_1 \rho_0/\xi_{\text{r}} \approx 84$ $\mu$m/s\\
Repulsion-induced slowdown slope&$v'_{\text{rep}} = -\zeta_0/\xi_{\text{t}}/(\pi R^2) \approx -68$ $\mu$m/s\\
Repulsion-induced diffusivity slope&$D'_{\text{rep}} = \zeta_1/\xi_{\text{t}}/(\pi R^2) \approx 3970$ $\mu$m$^2$/s\\
Torque-induced speed slope&$v'_{\text{tor}} = \tau_1/\xi_{\text{r}}/(\pi R^2) \approx 561$ $\mu$m/s
\end{tabular}
\bfcaption{Experimental estimates of parameters}{ The first part of the table lists estimates for parameters that are measured quantities. See \cref{electrodynamics} for the definition of the dipole factors. The second part of the table lists the formulae and estimates for parameters that are not directly measured but derived from other parameters. The third part of the table lists the formulae and values of parameters obtained from the measured pair distribution function (\cref{Fig gr}, see text). The last six quantities follow from the first three. The slope quantities at the end of the table, indicated with a prime, are defined here with respect to the area fraction $\phi$, not the particle concentration $\rho$.} \label{t parameters}
\end{center}
\end{table*}

Some of the estimates in the first part of \cref{t parameters} do not come from direct measurements in our experimental system. In particular, we estimate the viscosity $\eta$ and the relative dielectric permittivity $\epsilon_r$ of the solvent to be those of water at room temperature. Respectively, we estimate the electrostatic screening length $\lambda$ to be given by the distance between the electrodes in the experimental setup. Finally, the dielectric permittivity of vacuum, $\epsilon_0$, is a fundamental constant.

The rest of estimates in the first part of \cref{t parameters} are obtained by direct measurements in our experimental system. In particular, rotational fluctuations stem mainly from imperfections of the electrode surface, which slightly redirect particle motion over short time scales. Hence, the rotational diffusion coefficient $D_{\text{r}}$ needs not be related to the rotational friction coefficient $\xi_{\text{r}}$, which is mainly due to viscous drag from the solvent. From tracks of single-particle orientation at low particle densities, we measure $D_{\text{r}} \approx 0.15$ s$^{-1}$ (\cref{Fig speed-rotation}). This value reflects the non-thermal origin of rotational fluctuations; it is larger than the thermal value derived from the rotational Stokes-Einstein relation $D^{\text{SE}}_{\text{r}} = k_BT/\xi_{\text{r}}$, with $\xi_{\text{r}} = 8\pi \eta R^3$ for a spherical particle of radius $R$, which gives $D_{\text{r}}^{\text{SE}} \approx 0.05$ s$^{-1}$.

In the second part of \cref{t parameters}, we estimate the bare translational diffusion coefficient $D_{\text{t}}$ of our particles via the translational Stokes-Einstein relation $D_{\text{t}} = K_BT/\xi_{\text{t}}$, with $\xi_{\text{t}} = 6\pi \eta R$ the viscous drag coefficient of a sphere of radius $R$. We obtain a bare diffusion coefficient $D_{\text{t}} \approx 0.16$ $\mu$m$^2$, which is much smaller than the active contribution $v_0^2/(2D_{\text{r}})$ to the long-time diffusion coefficient of a self-propelled particle.

Finally, in the third part of \cref{t parameters}, we provide estimates for the coefficients $\zeta_0$, $\zeta_1$, and $\tau_1$, which characterize the different contributions to the collective interaction force and torque. To estimate these coefficients, we numerically evaluated the integrals in \cref{eq coefficients} using the experimentally measured pair correlation function $g(r,\phi)$ (\cref{Fig gr}). For convenience, we also provide estimates for other interaction coefficients directly derived from $\zeta_0$, $\zeta_1$, and $\tau_1$ (see \cref{eq interaction-functionals,eq slopes}).


\end{document}